\documentclass[]{aa}
\usepackage[varg]{txfonts}
\usepackage{natbib}
\bibpunct{(}{)}{;}{a}{}{,}\rm 
\usepackage{color}
\usepackage{graphics}

\begin{document}
\title{On the proper use of the Schwarzschild and Ledoux criteria in stellar evolution computations}
\author{M. Gabriel\inst{\ref{inst1}} \and A. Noels\inst{\ref{inst1}} \and J. Montalb\'an\inst{\ref{inst1}} \and A. Miglio\inst{\ref{inst2}}}
\institute{Institut d'Astrophysique et de G\'eophysique, Li\`ege University, All\'ee du 6 Ao\^{u}t, 17, B-4000 Li\`ege, Belgium \\\email{mgabriel@ulg.ac.be -- Arlette.Noels@ulg.ac.be -- j.montalban@ulg.ac.be}\label{inst1} \and School of Physics and Astronomy, University of Birmingham, Birmingham, B15 2TT, UK \\ \email{miglioa@bison.ph.bham.ac.uk}\label{inst2}}

\date{Received 16 January 2014 / Accepted}
\abstract{The era of detailed asteroseismic analyses opened by space missions such as CoRoT and \textit{Kepler} has highlighted the need for stellar models devoid of numerical inaccuracies, in order to be able to diagnose which physical aspects are being ignored or poorly treated in standard stellar modeling. We tackle here the important problem of fixing convective zones boundaries in the frame of the local mixing length theory. First we show that the only correct way to locate a convective zone boundary is to find, at each iteration step, through interpolations or extrapolations from points \textit{within the convective zone}, the mass where the radiative luminosity is equal to the total one. We then discuss two misuses of the boundary condition and the way they affect stellar modeling and stellar evolution. The first one consists in applying the neutrality condition for convective instability on the \textit{radiative} side of the convective boundary. The second way of misusing the boundary condition comes from the process of fixing the convective boundary through the search for a change of sign of a possibly \textit{discontinuous} function. We show that these misuses can lead to completely wrong estimates of convective core sizes with important consequences for the following evolutionary phases. We point out the advantages of using a double mesh point at each convective zone boundaries. The specific problem of a convective shell is discussed and some remarks concerning overshooting are given.}
\titlerunning{On the proper use of convection criteria} \rm 
\maketitle

\section{Introduction}
It appears that there is a misunderstanding concerning which criterion should be used to fix the boundaries of convective zones and also on which numerical procedure should be used to find them. The numerous methods tested recently by \citet{Paxton13} clearly show that stellar evolution results are very sensitive to the choice which is made. The answer to these questions cannot be found from comparing numerical results obtained with different assumptions in stellar evolution codes since other uncertainties affect stellar modeling. 

Therefore we think it is useful to recall some very basic physical facts which allow us to discuss the problem correctly and to bring a theoretical answer to the questions presently discussed. The extent of convective cores in stars is indeed crucial to fix the time frame of their evolution and is of particular importance for stellar and galactic evolution. Moreover, since models are often used to interpret seismic data, it is important that they be free from any numerical inaccuracy caused by an incorrect positioning of the convective zones boundaries so that any discrepancy between observation and theory may be attributed to inaccuracies in the physics. 

It is interesting to point out that when comparing stellar models obtained with different codes, as was done in the frame of the `CoRoT Evolution and Seismic Tools Activity'' (http://www.astro.up.pt/esta/) of the \textit{Seismology Working Group} of the CoRoT Mission (http://corot.oamp.fr/), the main differences are found in the location of convective zone boundaries and in layers close to them \citep{Lebreton08}. They are especially large in low mass stars models where the convective core mass increases during a large fraction of core hydrogen burning (see their Fig. 9). 

\rm For a long time the local mixing length theory (LMLT) was the only theory available for model computation but since a few decades, progresses have been made along two lines. The first one came with the progressive increase of computing power. However numerical simulations will always be limited to domains of Rayleigh and Prandtl numbers very different from these encountered in stars, which sometimes makes the results difficult to extend to stellar conditions. So far, no general model replacing LMLT has been inferred from numerical simulations but very interesting results have nevertheless been obtained, for instance concerning overshooting (see Sect. \ref{Over}). The second one came from the development of better theories. This started in 1981 with the first work of Xiong who went on improving his theory, now with several collaborators \citep{Xiong81, Xiong85, Xiong86, Xiong89, Xiong97, Xiong98a, Xiong98b, Xiong00}. The Chinese group produced a few evolutionary sequences (see Sect. \ref{Over}) but it is obvious that they have to solve a very unstable system with stiff equations and that model computation has become a much harder problem. 

Canuto has also worked out his own theory since 1992 (\citet[][and references therein]{Canuto99b}, \citet{Canuto00, Canuto11a, Canuto11b, Canuto11c, Canuto11d}) but so far no code has included his formalism. Though in order to progress, it would be highly desirable that people developing evolution codes introduce some better theory than the LMLT, we have to notice that this has not been done so far and worse, that some codes use a wrong numerical scheme to locate the boundaries of convective zones. 

In this paper we use the term semi-convective to refer to layers where the temperature gradient is in between the Schwarzschild and the Ledoux gradient. This does not imply that we consider that they are necessarily partially mixed; on the contrary, we consider that practically all local treatments of the problem may not be trusted. \citet{Canuto99b} has provided the equations which in principle allow us to derive the non-local solution in each situation but again no one has obtained a general non-local solution. The problem of the semi-convective zone structure  will not be discussed here. The purpose of this paper is very limited; we just want to recall what is the correct condition at the boundaries of convective regions for single non-rotating stars in the frame of the LMLT, and what is the only correct way to implement it. \rm

In Sect. \ref{Def} we recall the physical aspects and the definition of a convection zone and a convective boundary as well as the only correct method to find the position of that boundary which was already given by \citet{Biermann32} in the frame of the LMLT. 
Sect. \ref{Misuse1} discusses the consequence of a first misuse of the boundary condition, which is to apply the criterion fixing the neutrality condition on the radiative side of the boundary instead of its correct use on the convective side of the boundary. In Sect. \ref{Misuse2} another way of computing incorrect extents of convective regions is presented. In Sect. \ref{Models}, to illustrate the points discussed in the previous section, we have computed main sequence and core helium burning models with CL\'ES (Code Li\'egeois d'\'Evolution Stellaire) \citep{Scuflaire08} with different implementation of the algorithm used to fix the convective zones boundaries. Sect. \ref{Double} discusses the advantages of introducing a double mesh point at convective zones boundaries. Sect. \ref{Shell} raises the problems related to the occurrence of a convective shell in an otherwise radiative zone. A few remarks concerning overshooting are given in Sect. \ref{Over} and conclusions are drawn in Sect. \ref{Conclusions}.

\section{Definition of a convective zone boundary and of the correct boundary condition} \label{Def}
For any physical problem every basic mathematical equation must be the translation of a physical concept. Therefore, the definition of a convective zone boundary must be given out of a physical idea of what this boundary is. Only afterwards will it be possible to express it in terms of a mathematical formula.

A convection zone is a region where a fraction of the energy is carried by up and down motions of matter. Therefore the natural way to define the surface of a convection zone is to say it is the surface where the radial component $V_r$ of the velocity of the convective motions goes to zero. This definition was implicitly adopted from the very beginning; it was also given explicitly by \citet[]{Roxburgh78}. This definition is completely general and using it, the convection zone includes what is called the overshooting region. It says nothing on the temperature gradient at that point; a condition such that the radiative temperature gradient is equal to the adiabatic value comes in only in some convection theories and particularly in the local mixing length theory (LMLT).  

\rm In real stars this surface is very complex and therefore non spherical and varying with time. However, in model computation, we have to assume that it is spherical and varying only on the time scale of stellar evolution. This means that we have to search for an approximate solution of the equations for convection, which leaves out some phenomena which may be important, such as wave generation. It is well known that convection generates modes responsible for the solar-like oscillations in low mass stars. Convection can also generate gravity waves in overshooting regions, which can carry energy and angular momentum. In models computation, we also have to assume that convective zones are chemically homogeneous. A way to escape this approximation could perhaps be found if a solution of Cantuo's equations could be obtained which filters out the short time scales and keeps only those comparable with that of stellar evolution. (Theories such as those of \citet{Deng96} or of \citet{Ventura98}, which contain enough parameters to allow a large variety of results, may not provide a satisfactory solution.) Again, in particular when a convective boundary expands in the adjacent radiative one, this is wrong in the outermost part of the overshooting region where convection becomes so weak that the mixing time scale becomes of the order or larger than the evolution time scale. In these layers the chemical composition varies from that of the chemically homogeneous convective region to that of the radiative neighboring layers. This means that the density will then be continuous everywhere, as well as the opacity and the temperature gradient, and that most of the problems discussed in this paper will disappear (but a better theory than the LMLT is also required). According to \citet{Canuto99a} this layer is very narrow, so that a fine zoning is required to take it properly into  account and also that the discontinuity introduced when assuming chemically homogeneous  convective zones is a pretty good approximation. \rm 

The basic equations for stationary convection of interest here (\textit{i.e.} the \rm Eulerian \rm time derivative of any variable cancels out) in absence of any other velocity field, give \citep[see][]{Ledoux58, Gabriel96, Grigahcene04}
\begin{equation}
\overrightarrow{\nabla} \; \cdot \; \rho \overrightarrow{V} = 0 
\end{equation}
where $\overrightarrow{V}$ is the convection velocity.
This equation implies
\begin{equation}
\overline{\rho \overrightarrow{V}} = 0 \;\;. \label{rhoV}
\end{equation}
The overbar indicates the average over a spherical surface, which is large compared to the convective characteristic lengths. Eq. (\ref{rhoV}) simply means that there is no net flux of matter through a large spherical  surface. The equation of thermal energy conservation writes
\begin{equation}
\overrightarrow{\nabla} \; \cdot \; \left (\overrightarrow{F_R} + \overrightarrow{F_C} + \overrightarrow{F_K} \right ) = \overline{\rho \epsilon} - \frac{dU}{dt} + \frac{P}{\rho^2} \frac{d \rho}{dt}
\end{equation}
where $\epsilon$ is the nuclear energy generation rate and $U$ is the internal energy per unit mass of the mixture of gas and radiation. $\overrightarrow{F_R} $, $\overrightarrow{F_C}$ and $\overrightarrow{F_K}$ are respectively the radiative flux, the convective flux and the flux of kinetic energy of convection, with
\begin{equation}
\overrightarrow{F_C} = \overline{\left (\frac{P}{\rho} + U \right ) \rho \overrightarrow{V}} \;\;,\label{Fc1}
\end{equation}
 and
\begin{equation}
\overrightarrow{F_K} = \frac{1}{2}\overline{V^2 \rho \overrightarrow{V}} \;\;. \label{Fk}
\end{equation}
These equations are very basic ones and are independent of any theory of convection. The only approximation done in Eq. (\ref{Fc1}) is to assume that the gas and the radiation are in thermal equilibrium (more general equations are given in \citet{Gabriel96}). 

We shall now assume that the convective pressure fluctuations are much \rm smaller \rm than the temperature and density ones so that they can be neglected. This approximation is practically always done in stellar convection theories and is valid as long as the convection velocity is small compared to the sound speed. 
We also assume that the density and temperature fluctuations are small enough so that we can neglect their powers larger than one. It then follows that
\begin{equation}
\overrightarrow{F_C} = \rho T \overline{\left [\Delta S \overrightarrow{V}\right ]} = C_P \rho T \overline{\left [\frac{\Delta T}{T} \overrightarrow{V}\right ]} \;\;.
\label{Fc2}
\end{equation}
The fluctuation of any variable $f$ is $\Delta f$. Notice that to obtain equation \ref{Fc2} from Eq. (\ref{Fc1}) it is important to consider $\rho \overrightarrow{V}$ as a single variable in order to take properly Eq. (\ref{rhoV}) into account.
Eq. (\ref{Fc2}) is the relation usually used for the convective flux. The flux $\overrightarrow{F_K}$ is nearly always neglected in stellar model computations because it cancels out in LMLT. 

\rm In spherically symmetric stars $\overrightarrow{F_C}$ is a radial vector and from our definition of a convection zone boundary, $V_r$ is equal to zero everywhere on the spherical surface boundary. It follows from Eqs. \ref{Fc1} and \ref{Fk}, that the convective flux $\overrightarrow{F_C}$ and the flux of mechanical energy of convection $\overrightarrow{F_K}$ are equal to zero on that surface and also that the radiation carries out all the energy, \textit{i.e.}
\begin{equation} 
\overrightarrow{F_R} =\overrightarrow{F}  \;\; \textrm{and} \;\; V_r = 0 \;\;.
\label{cond}
\end{equation}

In the frame of the LMLT, it can be replaced by $L_R = L$  because $\overrightarrow{F_K}$  is neglected and consequently there is no point where  $\overrightarrow{F_C} = - \overrightarrow{F_K}$ as in better theories. \citet{Canuto99a, Canuto11d} has suggested to replace the boundary condition (\ref{cond}) by the condition that the helium or hydrogen concentration flux is equal to zero, \textit{i.e.} that  $\overline{\rho Y \overrightarrow{V}} = 0$ (as usual $Y$ is the helium mass abundance). This will indeed provide a more precise location of the boundary as convection will be very inefficient in the outer overshooting zone so that  $\overrightarrow{F_C}$ and $\overrightarrow{F_K}$ will be practically zero over an extended region  while the chemical composition is expected to vary quite rapidly near the surface. Unfortunately with a simple minded theory such as LMLT, convective region and the \textit{ad hoc} added overshooting zones have to be chemically homogeneous. \rm 

The condition (\ref{cond}) can also be obtained from the following approach. The equation of radiative transfer holds everywhere in a star and the only terms which can be discontinuous are the source function and the opacity coefficient. As a result only the gradient of the radiative intensity can be discontinuous. It follows from their definitions that the radiation energy per unit volume, the radiation flux and the radiation pressure are all continuous throughout a star \citep[see for instance][p. 39]{Schwarzschild58}

Obviously the boundary condition $V_r=0$ or $\overrightarrow{F_C} = \overrightarrow{F_K} =0$ is meaningful only in a convection zone and consequently condition (\ref{cond}) must be applied \textit{on the convective side of the boundary and not on the radiative side}. 

In stellar evolution the local B\"{o}hm-Vitense theory is commonly used to find the structure and to fix the position of the boundaries of a convection zone. Therefore in this paper we shall only consider models computed in the frame of a local theory of convection, such as LMLT, and without overshooting since it is a non local phenomenon. So far no theory reliably predicts the properties of the overshooting region. Accordingly its extent is just a free parameter often taken as a fraction of the local pressure scale height while the assumed temperature gradient in this region differs from one code to the other. However, short remarks concerning overshooting will be given in Sect. \ref{Over}.

Already \citet{Biermann32} showed that convective regions may exist in \rm the stellar interior \rm and that the temperature gradient is adiabatic there.  More precisely, using LMLT, he showed that $(\nabla - \nabla_{ad})$ is very small ($\nabla = \frac{d\ln T}{d\ln P}$ and $\nabla_{ad} = (\frac{d\ln T}{d\ln P})_S$) and also that $V_r^2 \propto (\nabla - \nabla_{ad})$. Therefore at the boundaries of a convection zone the temperature gradient is adiabatic. Since $\overrightarrow{F_C} = \overrightarrow{F_K} = 0$ there, it follows from \rm condition (\ref{cond}) \rm that
\begin{equation}
\nabla_R = \nabla_{ad} \;\; .\label{Sch}
\end{equation}
Clearly this condition, which is called the \textit{Schwarzschild criterion}, must be satisfied on the convective side of the boundary, (just as the conditions $\overrightarrow{F_C} = \overrightarrow{F_K} =0$ and $\overrightarrow{F_R} = \overrightarrow{F}$). This is the conclusion already obtained by \citet[][see Eq. (30)]{Biermann32}. On the radiative side of the boundary, we must have 
\begin{equation}
\nabla_R \leq \nabla_{ad} \;\; .
\end{equation}
The equality holds when there is no discontinuity in the chemical composition since then all the variables $P, T, \rho$ and $L$ are continuous. The inequality holds when there is a discontinuity of the chemical composition as then both the opacity and the density are discontinuous. \textit{Condition (\ref{Sch}) gives the boundaries of a convective zone only in the frame of local convection theories}.

When convection takes place in superficial layers with low temperatures and densities, a \rm non-negligible \rm departure of the temperature gradient from the adiabatic one is required but as long as a local theory such as \rm B\"{o}hm-Vitense's \rm LMLT is used, it is found that the condition $V_r=0$ or $\overrightarrow{F_R} = \overrightarrow{F}$ still leads to $\nabla_R = \nabla_{ad}$ at the surface of the convective envelope. Consequently and since stellar models are usually computed with local convection theories, we shall use, in the forthcoming discussion, the definition of a convective zone boundary given by LMLT, \textit{i. e.} $V_r=0$ or $\overrightarrow{F_R} = \overrightarrow{F}$ or $\nabla_R = \nabla_{ad}$ \textit{taken on the convective side of the boundary}. This implies that since during the iteration process of stellar modeling the assumed convection zone boundaries are not the right ones, \textit{their position must be improved through extrapolations or interpolations of $L_R$ from points in the convective zones in order to find the mass where $L_R = L$ and $\nabla_R = \nabla_{ad}$.} This is the way Henyey's code worked \citep[][bottom of page 309]{Henyey64}. \footnote{In 1968, after one of us (MG) was in Berkeley on a postdoctoral position, L. Henyey gave us his code. Of course we made it evolve with time, but we always kept its main characteristics, \textit{i.e.} the same variables, the same way to handle convective zone boundaries and the introduction of double mesh points. We used it until 1999.}

It is also interesting to notice that the boundary \rm condition (\ref{Sch}) \rm was correctly applied at least up to 1958 when models were computed using the fitting technique. Most computed models were two zones models with a discontinuity of molecular weight between the convective core and the radiative envelope. Such models were computed for instance by \citet{Ledoux47} and \citet{Oke52}. Models with a $\mu$-gradient region in between were first computed by \citet{Tayler54,Tayler56} and then by \citet{Kushwaha57}. All these authors applied condition (\ref{Sch}) on the convective side of the boundary, without justification since it was evident to them. We only find a detailed justification in Schwarzschild's famous book \citep[][pp. 167,168]{Schwarzschild58}. He clearly explains why condition (\ref{Sch}) must be applied on the convective side of the core boundary and not on the radiative one. 

On the other hand the Ledoux criterion was likely introduced in stellar modeling only after the discovery of main sequence models with a semi-convective zone because it allows the computation of models showing up these layers though ignoring what their real structure is. On the opposite, with the Schwarzschild criterion, these layers are considered as convective. This leads to convergence problems which disappear only when a theory which specifies the temperature gradient and the way to compute the chemical composition in these semi-convective regions is adopted. 

Since the correct boundary condition must be obtained from points in the convective zone only, any use of points in the radiative one may lead to an incorrect positioning of the convective zone boundary. When points located in the radiative zone are nevertheless used it is possible to apply either the Schwarzschild \rm criterion with $y =\nabla_R - \nabla_{ad}$ \rm or the Ledoux criterion with $y = \nabla_R - \nabla_{Ldx}$ \rm where \rm

\begin{equation}
\begin{array}{lll}
\nabla_{Ldx} & = & \nabla_{ad} - \left (\frac{\partial \ln P}{\partial \ln T} \right )^{-1}_{\rho,X_i}  \sum_i  \left ( \frac{\partial \ln P}{\partial \ln X_i} \right )_{\rho,T,X_j} \frac{d\ln X_i}{d\ln P}   \label{Ldx} \\
& = & \nabla_{ad} + \left ( \frac{\beta}{4 - 3 \beta} \right ) \frac{d\ln  \mu}{d\ln  P} 
\end{array}
\end{equation}
\rm and \rm the last equality holds for the simple equation of state $ P = \mathcal{R} \rho T / \beta \mu$ \rm with \rm $\beta = P_g/P$.

Firstly a question might be asked. Which one of the criteria must be preferred, Schwarzschild's or Ledoux's? The question is of course meaningless since only \rm points within the convective region \rm may be used and that both criteria are identical there. The use of points in the radiative zone raises problems when the function $y$ is discontinuous at the convective zone boundary. With the Schwarzschild criterion this happens only when the chemical composition is discontinuous there. With the Ledoux criterion, $y$ is discontinuous when the chemical composition is discontinuous at the convection boundary and also when there is just a $\mu$-gradient in the adjacent radiative layers. Therefore we may say that the Ledoux criterion more often leads to problems than the Schwarzschild one in models computation with the LMLT. \rm This however is only meaningful for numerical techniques but has no physical relevance and, as just said, the question is meaningless. \rm 

One might also wonder where these misuses of the boundary condition come from. Clearly they come from the use of Eq. (\ref{Sch}) instead of the most fundamental one given by Eq. (\ref{cond}). Since in the radiative zone $L_R=L$ by definition, while in the convective one $L_R<L$ , it is obvious that the only possible way to find the mass point of the convective region where radiation is finally able to carry out all the luminosity is through interpolation or extrapolation from points in the convective region, and that points in the radiative layers may be of no help for that purpose. When the Schwarzschild or the Ledoux criterion is used, the variable $y$ defined above varies everywhere in the star and is nowhere constant, but one has to remember that it is just another way to write the original condition (\ref{cond}). When this is forgotten, one can be tempted to use points in the radiative zone. This works well when $y$ is continuous at the interface between the two regions but, as will be shown below, when $y$ is discontinuous there, the use of one or several radiative points will most often lead to an incorrect positioning of the boundary. \textit{This shows up very clearly when $L-L_R$ or $\nabla_R-\nabla_{ad}$ is plotted. It is then seen that these quantities do not cancel on the convective side of the boundary.} 

Secondly there are two ways of misusing the boundary condition. In the first one, the zero of $y$ is obtained by interpolation or extrapolation from points in the radiative region; this is discussed in Sect. \ref{Misuse1}. In the second one a change of sign of $y$ between two consecutive mesh points is searched for; these two points may be both radiative or both convective or one may be radiative and the other convective but $y$ is then always interpolated between these two points. This is discussed in Sect. \ref{Misuse2} and illustrated in Figs. \ref{bv_16M}, \ref{bv_1.5M} and \ref{bv_8M} of Sect. \ref{Models}.

The ratio of the radiative luminosities on both side of the boundary is given by
\begin{equation}
\frac{L_{R,i}}{L_{R,e}} = \frac{\nabla_i}{\nabla_e} \frac{\kappa_e}{\kappa_i} \label{Lratio}
\end{equation}
where $\kappa$ is the opacity coefficient and the indices $e$ and $i$ refer respectively to the outer and inner sides of the convective boundary. Any misuse of the boundary condition leads to the violation of either Eq. (\ref{cond}), $L_{R,i} = L$, or of the obvious one, $L_{R,e} = L$.

\section{On the first way to misuse the boundary condition} \label{Misuse1}

\subsection{With the Ledoux criterion} \label{Mis1Ldx}
One way is to use the Ledoux criterion for convective instability ($\nabla_R > \nabla_{Ldx}$) and to require that it predicts neutrality with $\nabla_R - \nabla_{Ldx} =0$ on the radiative side of the boundary.    
We have to distinguish several possibilities occurring at a convective zone boundary.

\subsubsection{The chemical composition is discontinuous but there is no gradient of molecular weight in the radiative zone}
The Ledoux criterion is then identical to the Schwarzschild one. By hypothesis we take in this subsection $\nabla_e = \nabla_R = \nabla_{Ldx} = \nabla_{ad}$. Moreover in a deep convective zone, the temperature gradient is adiabatic and thus $\nabla_i = \nabla_{ad}$. We obviously have $L_{R,e} = L$ and Eq. (\ref{Lratio}) shows that except in the unphysical situation where the opacity is independent of the chemical composition, we obtain $L_{R,i} \neq L$. We must consider two different cases:

\begin{enumerate}
\item $\kappa_e < \kappa_i$

Then $L_{R,i} < L$ and the convective luminosity must be positive, which implies a positive convective velocity. This means that convective motions will extend further out than the assumed boundary, which then is not the right one since $V_r \neq 0$. This situation was discussed by \citet[][pp. 167,168]{Schwarzschild58}. He explicitly showed that the boundary condition must be applied on the convective side. 

On the contrary, if the condition $\nabla_R = \nabla_{ad}$ (or $L_{R,i} = L$) is applied  on the convective side, a subadiabatic temperature gradient is sufficient for radiation to carry out all the luminosity in the radiative region. 

\item $\kappa_e > \kappa_i$

Then $L_{R,i} > L$ which means that a subadiabatic temperature gradient is large enough to have radiation carry out all the luminosity. As a result, in the outer layers of the convection zone, $F_C < 0$ and consequently $\nabla < \nabla_{ad}$. This means that they are stable and the assumed boundary must be moved inwards.  

If the condition $\nabla_R = \nabla_{ad}$ is correctly applied on the convective side of the boundary, then $L_{R,i} = L > L_{R,e}$ and layers located on the radiative side are convectively unstable. This is the difficulty met by \citet{Harm58} when they studied the evolution of massive main sequence stars where electron scattering is the main source of opacity. \rm This led them to introduce semi-convection into models of these stars. \rm
\end{enumerate}

\subsubsection{The chemical composition is continuous but the abundance gradients of some abundant elements are discontinuous}
This situation is encountered at the surface of the convective core of intermediate mass stars during core hydrogen burning phases. 
Since in all the situations encountered in stellar evolution the chemical composition gradient terms in the Ledoux criterion have a stabilizing influence, one has $\nabla_e = \nabla_R  = \nabla_{Ldx} > \nabla_{ad}$ on the radiative side. Relation \ref{Lratio} is now written with $\kappa_e = \kappa_i$ and thus $\nabla_i  = \nabla_{ad} < \nabla_R$. Therefore   
\begin{equation}
\frac{L_{R,i}}{L_{R,e}} = \frac{L_{R,i}}{L} < 1
\end{equation}
which means that $\overrightarrow{F_C} > 0$ and $V_r > 0$ at the boundary of the convection zone.This implies that convection will extend further out than assumed. 

Had we applied the condition  $\nabla_R = \nabla_{ad}$ on the convective side of the boundary, then $\nabla_e = \nabla_{ad}$  and the chemical composition gradient terms forces the Ledoux criterion to predict stability on the radiative side of the boundary in all known situations.

\subsubsection{The chemical composition and its gradient are discontinuous} 
Here Eq. (\ref{Lratio}) gives
\begin{equation}
\frac{L_{R,i}}{L} =  \frac{\kappa_e}{\kappa_i} \frac{\nabla_{ad}}{\nabla_{Ldx}} < \frac{\kappa_e}{\kappa_i}   \;\;. \label{Ldx3}
\end{equation}
Again we have to distinguish two possibilities:

\begin{enumerate}
\item $\kappa_e < \kappa_i$

Then $L_{R,i} < L$ on the convective side which means that convective motions do not vanish there and that the convection zone must be extended.

\item $\kappa_e > \kappa_i$

Then one might find $L_{R,i} = L$  but only for a very special combination of $\kappa_i / \kappa_e$ and of the gradient of molecular weight. In most cases it will be found that $L_{R,i} \neq L$ and that the convective boundary must be moved in one or the other direction. This situation is met in MS low mass stars with a small growing convective core and nuclear reactions outside this convective core. 

\end{enumerate}

Had we used the correct boundary condition, we would have found that  $\nabla_e = \nabla_{ad} \kappa_e / \kappa_i$ and the outer side of the boundary is stable provided that
\begin{equation}
\nabla_e = \nabla_{ad} \frac{\kappa_e}{\kappa_i} < \nabla_{ad} - \left (\frac{\partial \ln P}{\partial \ln T} \right )^{-1}_{\rho,X_i}  \sum_i  \left ( \frac{\partial \ln P}{\partial \ln X_i} \right )_{\rho,T,X_j} \frac{d\ln X_i}{d\ln P}  \frac{d\ln  \mu}{d\ln  P} \;\;.
\end{equation}

If $\kappa_e < \kappa_i$ there is no problem indeed. If $\kappa_e > \kappa_i$ it is at first found that $\nabla_i = \nabla_{ad} < \nabla_e = \nabla_R < \nabla_{Ldx}$; such layers are often considered as semi-convective. However, when the ratio $\kappa_e / \kappa_i$ is large enough, we find that the radiative side of the boundary is convectively unstable towards the Ledoux criterion leading again to a difficulty but we do not know of any case where such a situation is met. 

To our knowledge this method with the Ledoux criterion was never used in all cases discussed here in Sect. \ref{Mis1Ldx} because it leads to a much too fast convective core decrease due to the strong stabilizing influence of the $\mu$-gradient term during core hydrogen burning. One can expect an even faster shrinking than that found in \citet{Paxton13} in their Fig. 13 and a Hertzsprung-Russell track still less compatible with observations than what they found in their Fig. 14.

\subsection{With the Schwarzschild criterion} \label{Schwarzschild}
Another way to misuse the boundary condition is to apply the Schwarzschild criterion (Eq. (\ref{Sch})) on the radiative side of a boundary. We have to distinguish the same 3 cases as above:

\begin{enumerate}
\item For the first one Schwarzschild and Ledoux criteria are identical and the same conclusions as above are reached. 
\item In the second case, as the chemical composition is continuous so are the opacity, the radiative luminosity,  $\nabla_R$ and $\nabla_{ad}$. As a result the boundary condition leads to the same conclusion as the right one. 
\item In the third case, the use of the Schwarzschild criterion leads to the same conclusions as in the first one. 
\end{enumerate}

This incorrect boundary was used by many researchers among these who, after 1960, wrote evolution codes using the finite difference method developed by \citet{Henyey54a,Henyey54b}. Since during hydrogen burning the convective core generally shrinks, there is no discontinuity in chemical composition there and this mistake has then no consequence. 

But later on when central helium burning was computed, it was found that the convective core expands and builds up a discontinuity in chemical composition. The mistake, which had important consequences on the models structure, was noticed and corrected by \citet{Castellani71a,Castellani71b} who rediscovered the justification given earlier by \citet{Schwarzschild58}. The correction of that mistake also led them to discover that the mass of the convective core increases much more than with the incorrect boundary condition. When it becomes large enough, the physical situation discussed in \citet{Ledoux47} applied to the models they were considering.  Following Ledoux, they introduced a semi-convective zone in some helium burning models. This example nicely illustrates the consequences of a mistake done perhaps consciously because it then had no consequences and was easier to implement than the correct one but which led to incorrect results later on when an initially unexpected situation was encountered.

Summarizing we repeat what was said at the beginning: \textit{the criterion fixing the boundaries of convective regions must always be checked on the convective side of the boundary and never on the radiative side.} This criterion is $V_r = 0$  or $\overrightarrow{F_R} = \overrightarrow{F}$, which, in the local mixing length theory, is given by the neutrality of the Schwarzschild criterion $\nabla_R = \nabla_{ad}$, equivalent to the Ledoux criterion there.

On the other hand the Ledoux criterion must \textit{only and necessarily} be used to answer the following question: In a region of varying mean molecular weight, given a layer supposedly in radiative equilibrium, when is the temperature gradient large enough for convective motions to start there?

\section{On the second way to misuse the boundary condition} \label{Misuse2}
Another way to misuse the convective zone boundary condition is found in many current codes. They scan the mesh points from the surface to the center (or the other way around) searching for a change of sign of either $(\nabla_R - \nabla_{Ldx})$ or $(\nabla_R - \nabla_{ad})$ and then they interpolate in order to find where it cancels. The procedure fails when the checked variable $y$ is discontinuous at the boundary of the convective zone. It is always so with $y =  (\nabla_R - \nabla_{Ldx})$ (except for homogeneous models), and only when the chemical composition is discontinuous with $y =  (\nabla_R - \nabla_{ad})$. The reason is very simple. If, at one step during the iteration process, $y$ changes sign between the two mesh points $j$ and $j+1$, then the convection zone boundary is assumed to be located between these two points. However, it is obviously not allowed to interpolate a function in an interval over which it has a discontinuity. This may have important consequences on the position of the convection zone boundary as we will now show. 

Let's consider the problem raised by a convective core, which is the most often encountered situation. The following discussion is also valid for the upper boundary of a convective shell and it is straightforward to transpose it for the bottom of a convective envelope or the lower boundary of a convective shell.

\subsection{Overview of the problem} \label{Overview}
We first present the problems discussed in the following with a few very schematic figures.  We consider a function $y(m)$ where $y$ is given either by $y = \nabla_R - \nabla_{ad}$ or by $y = \nabla_R - \nabla_{Ldx}$ and, as said above, we discuss here the problem of locating the mass at the convective core boundary, which is \textit{a priori} unknown. 

When $y$ is continuous throughout the model, the location of the point $y = 0$  is unambiguous. Problems only arise when $y$ is discontinuous at the location of the core boundary. First we consider the most commonly encountered situation: $y$ is smaller on the radiative side of the discontinuity that on the convective one. 

Since the transfer equation is not the same in convective and radiative zones, two different sets of differential equations are solved in these regions. As a result, it is possible to compute a relaxed model with, in our example, an arbitrary value for the convective core mass, that is to say a model such that $r, P, T$ and $L$ are continuous there and also such that the chemical composition varies according to the condition required by the sign of the time derivative of the convective core mass. \rm However, among all these models, only one will also fulfill the condition (\ref{cond}), $L_R = L$ or $\nabla_R = \nabla_{ad}$, on the convective side of the core boundary. In practice, we will deal with approximate models obtained after an iteration step and, in any reasoning done with them, we must hope that they are not too different from the relaxed one with the same core mass. \rm

\begin{figure}
\hspace*{-6mm}
\includegraphics[scale=0.34,angle=270]{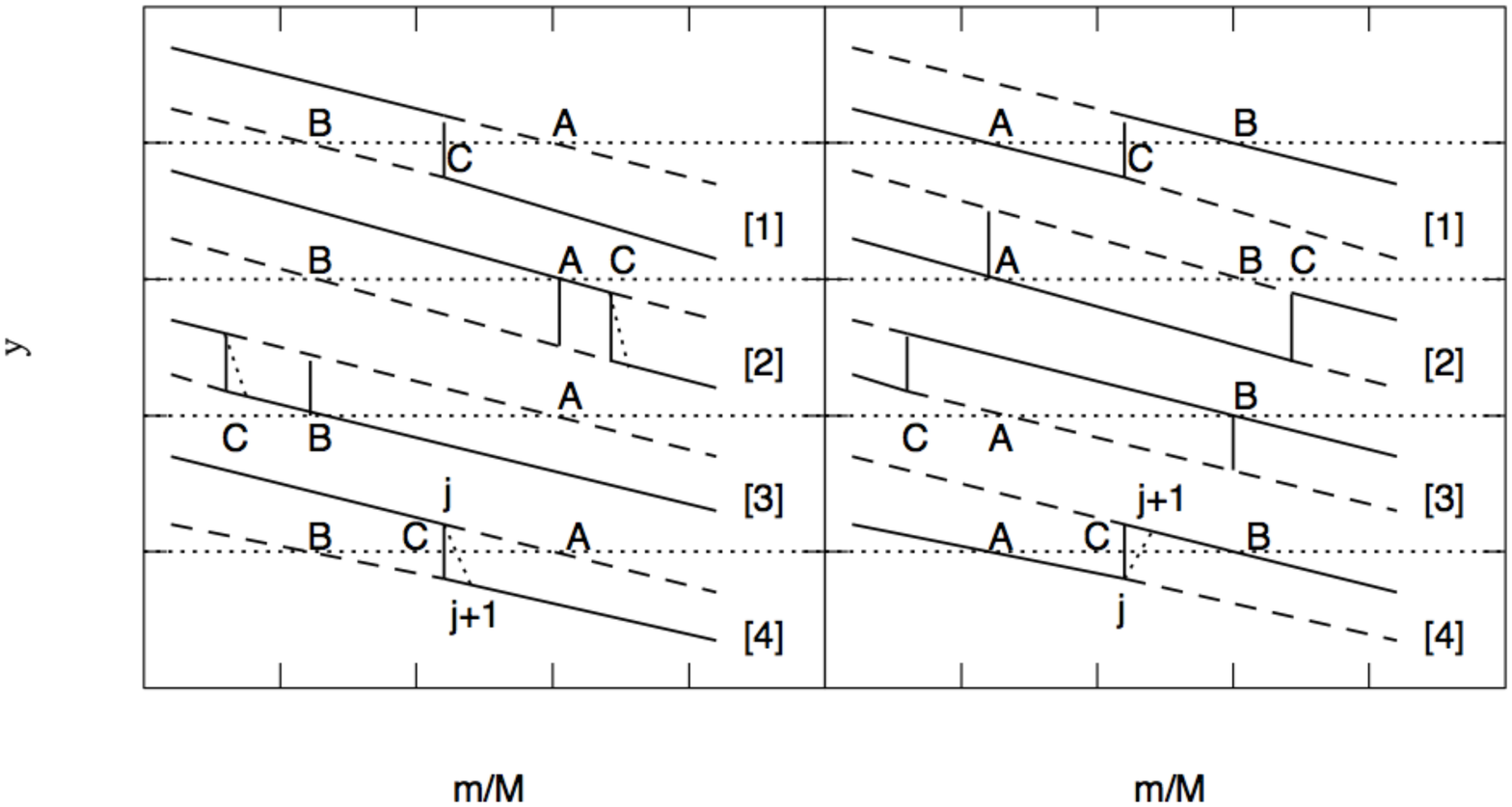} 
\caption{In the left (right) panel $y$ is smaller (larger) on the radiative side of the convective core boundary. Cases 1 to 4 stand for different situations. Case 1 illustrates two possible ways of fixing the core boundary. Cases 2 to 4 show the different solutions found when the algorithm searches for a change of sign of $y$ in the trial model}
\label{scheme}
\end{figure}

We assume that $y = Y_1$ for $m \leqslant m_C^{-}$, \textit{i.e.} in the convective core, and that  $y = Y_2$ for $m \geqslant m_C^+$, \textit{i.e.} in the radiative zone, and also that $Y_1$ and $Y_2$ are two continuous functions in the whole mass domain of interest. In the four cases illustrated in the left panel of Fig. \ref{scheme}, the upper curve is $Y_1$ and the lower one is $Y_2$ (they are shown partly as dashed lines). In the right panel, the positions of $Y_1$ and $Y_2$ are inverted. To simplify the problem, we assume that $Y_1$ and $Y_2$ do not change when the core mass varies. We consider that, as in Fig. \ref{scheme} (left panel, case 1), such a model has been computed with a core boundary tentatively placed at point $C$. The solution representing the model is given by the full line. 

If a new position of the core boundary is found through the extrapolation of the solution from the convective core and the new boundary is taken as the mass where $Y_1(m) = 0$ (point $A$), then the model computed with that boundary is the correct one as shown in Sect. \ref{Def}. 

Another procedure which also gives a unique solution is to obtain the core boundary from an extrapolation of the solution from the radiative zone and locating the core boundary at point $B$. This procedure is incorrect indeed as was shown in Sect. \ref{Misuse1}.

 The position of the core boundary found by these two methods is independent of the initial guessed position of that boundary as can be seen in cases 2 to 4 where interpolations may indeed be used in some cases. It is also independent of the mesh points distribution.

The most often used procedure searches where $y$ changes sign in the converged trial model (or in a given iteration). Let us briefly see what happens then. First we assume that the guessed core mass is too large, \textit{i.e.} $m_C > m_A$ (case 2). Again the full line represents the model and the interpolation of $y(m)$, which is done in the convective core of the current model, gives $y(m_A) =0$, the correct location of the convective core boundary.

Let us now assume that the initial choice of the core mass is such that $m_C < m_B$ (case 3), then the zero of $y$ located along the solution is at point $B$, which is not the right location of the core boundary. 

Let us finally consider the worst possibility, the initial core mass is such that $m_B < m_C < m_A$ (case 4). Then $y$ changes sign at the location of the discontinuity! This means that any trial mass $m_C$ between $m_A$ and $m_B$ will appear to be the solution. This procedure must be banished as it finds different solutions when the initially guessed core mass varies. 

More generally if during the iteration procedure the core mass does not always remain larger than the correct value, then the solution found will be incorrect. \rm Since there is a discontinuity of $y$ at the core boundary, a \textit{double mesh point}, \textit{i.e.} two points with the same mass but different values of the density and/or of the chemical composition gradient, should be introduced there. Unfortunately, \rm this is not generally the case and CESAM \citep{Morel08} is the only code we know which puts a double mesh point at the boundary of a convective zone when the density is discontinuous there. In the other codes the discontinuity is replaced by a segment connecting the upper side of the discontinuity, which is point $j$ \textit{i.e.} the last point in the convective core, to point $j+1$, the first point in the radiative zone. It is readily seen that if interpolations are done between these two points they will just bring asymptotically the core boundary to point $j+1$.

The right panel in Fig. \ref{scheme} shows the same problem when $y$ is larger on the radiative side than on the convective one of the discontinuity. The discussion of the different cases is the same as for the left figure and we will not repeat it. Let us just point out that  point $A$ still gives the correct position of the convective core boundary and also that now $m_A < m_B$. This situation will be further discussed at the end of Sect. \ref{ccgrow}.

\subsection{Analysis of the possible situations} \label{Analysis} 

At some step of the iteration process, the estimated mass of the convective core $m_C$ is either too small (see Fig. \ref{graph12}, left panel) or too large (see Fig. \ref{graph12}, right panel). The full lines in Fig. \ref{graph12} represent a schematic distribution of $y$ as a function of the fractional mass for a model without discontinuity (case 1), for a model with a small discontinuity (case 2) and for a model with a large discontinuity (case 3) at the boundary of the convective core. The dashed lines show the extrapolations of $y$ from the convective layers upwards and $A$, the point where it is equal to zero, gives the correct estimate of the convective boundary location. The dashed lines and the full lines in case 1 are only identical for homogeneous models. Notice that since most codes do not add a double mesh point at the core boundary, they replace the discontinuity by a fast variation of $y$ between $j$ and $j+1$, shown by the doted lines in Fig. \ref{graph12}.

\begin{itemize} 
\item{\textit{The estimated convective core is too large (Fig. \ref{graph12}, right panel)}} 
\begin{figure}
\hspace*{-6mm}
\includegraphics[scale=0.34,angle=270]{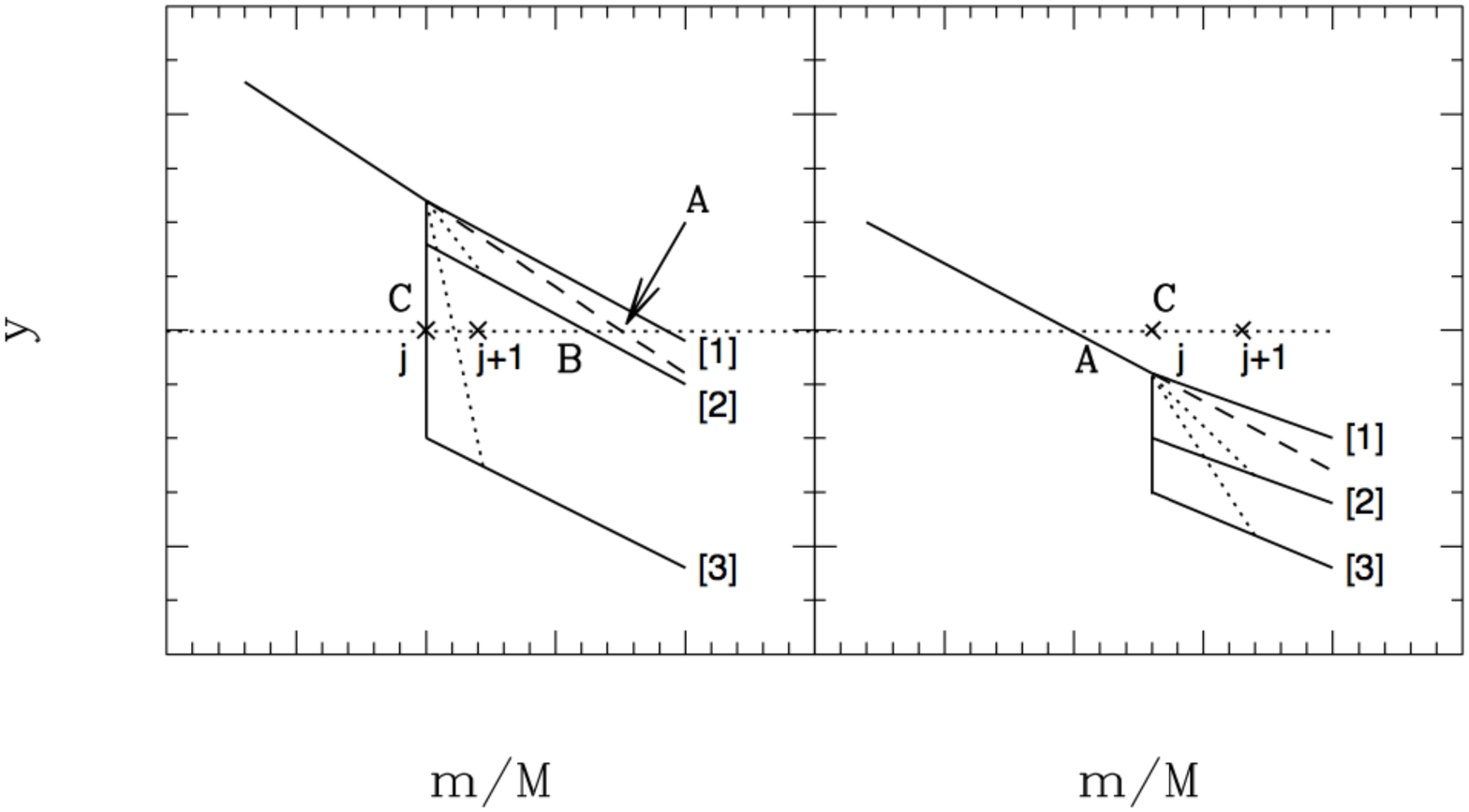} 
\caption{The left panel illustrates a situation such that the assumed core mass of a model (not necessarily converged) is too small. In the right panel the assumed core mass is too large. The full lines represent $y$ in the model while the dashed ones are extrapolations. Case 1 stands for a model without discontinuity at the core boundary when the Schwarzschild criterion is used. Case 2 represents situations such that $y$ has a small discontinuity. Case 3 stands for a larger discontinuity}
\label{graph12}
\end{figure}

The zero of $y$ is always in the convective core where  $y$ is continuous, the interpolation of $y(m)$ is allowed and it
leads to a significant improvement of the core boundary location since point $A$ gives its correct position. It is always so when $y$ is continuous (case 1) but in the two other cases, as soon as the location has been improved enough so that $y$ changes sign between points $j$ and $j+1$, $y$ is discontinuous in the interval and it is meaningless to try to improve its position through an interpolation of $y$ between these two points. This is the situation illustrated by case 3 in the left panel of Fig. \ref{graph12} (see discussion below) for which the correct convective boundary, given by point A, can fall outside the interval $[j,j+1]$ depending on the value of $y$ at point $j$ (see also Sect. \ref{Def} and \citet[][pp. 167,168]{Schwarzschild58}).
\vspace*{3mm}

\item{\textit{The estimated convective core is too small (Fig. \ref{graph12}, left panel)}}
In cases 1 and 2, $y$ is equal to $0$ inside an interval where $y$ is continuous and the interpolation will improve the location of the core boundary (which will be close to point A). The level of improvement nevertheless decreases as the discontinuity of $y$ increases. 
In case 1 a new core mass larger than $m_A$ is found and at the next iteration we are back to the situation discussed previously (right panel).

In case 2, if the discontinuity of $y$ is very small (even smaller than the one displayed for case 2 in the left panel of Fig. \ref{graph12}), the interpolation along the solution in the radiative zone can predict a value of $m_B > m_A$ and at the next iteration the core mass will be too large, which again will bring us back to the previous case (Fig. \ref{graph12}, right panel). But as the discontinuity becomes large enough compared to $y_j$ it will be found that $m_B < m_A$ and the algorithm will become unable to increase the core mass up to its correct value, which is close to point $A$. When this core mass is adopted at the next iteration, $y$ will be discontinuous at that point. As soon as the core boundary given by that algorithm has been located with the maximum accuracy it is able to achieve, \textit{i.e.} when $y$ changes sign between points $j$ and $j+1$ then case 2 evolves into case 3. 

In case 3 the discontinuity of $y$ occurs at the location of the assumed core boundary, supposed to be at point $j$, and the interpolation is thus meaningless (since with a correct physics there is a discontinuity inside that interval) and moreover completely fails to improve the boundary location which will remain embedded in the layer $[j,j+1]$ while the correct boundary, given by point A, can be located outside the interval $[j,j+1]$ depending on the value of $y$ at point $j$. If such a situation is encountered at the first iteration, the algorithm does not even allow any change in the core boundary location \citep[see for example][Fig. 15]{Paxton13}. This simply means that the algorithm does not work at all in case 3 and in case 2 as soon as the discontinuity in $y$ is large enough. Moreover the error made on the core mass may increase with the importance of the discontinuity since $y$ may become large on the convective side of the discontinuity.
\end{itemize}
\textit{When the estimated convective core is too small, the algorithm discussed here will generally predict an incorrect convective boundary while with a too large estimation, the final boundary will be underestimated if an oscillation in the core mass occurs during the iteration process.}

\section{Examples with real stellar models} \label{Models}
\subsection{Core hydrogen burning}
\subsubsection{The convective core shrinks during MS}

The method will converge toward the correct solution, provided the Schwarzschild criterion is used, since $y$ is continuous at the core boundary and as long as there is no semi-convective zone on top of the convective core. Then, as already pointed out, convergence problems arise, which can only be properly solved after doing some hypotheses on the structure of a semi-convective zone. These problems will be discussed in Sect. \ref{ccgrow}.

If the Ledoux criterion is used, then for the algorithm presented above to be valid in all three cases considered in Fig. \ref{graph12}, it is necessary to know beforehand the sign of the time derivative of the core mass and to always have a guessed core mass which is too large. Since in most MS stars the convective core mass decreases with time, a commonly used algorithm is to choose, for the zero order approximation core mass of the computed model, the same value as in the previous one. However, if during the iterations a guessed core mass is too small, the algorithm will predict an incorrect boundary and the extent of the convective core mass will be too small. As a result, its core hydrogen abundance decreases too quickly and this leads to a faster decrease of the core mass. This kind of problem with the guessed core mass during the iteration process is likely to occur several times during the main sequence phases, which, as a result, will be too short. This obviously has an important impact on the later stages of the star evolution. 

\begin{figure}
\hspace*{-6mm}
\includegraphics[scale=0.35,angle=270]{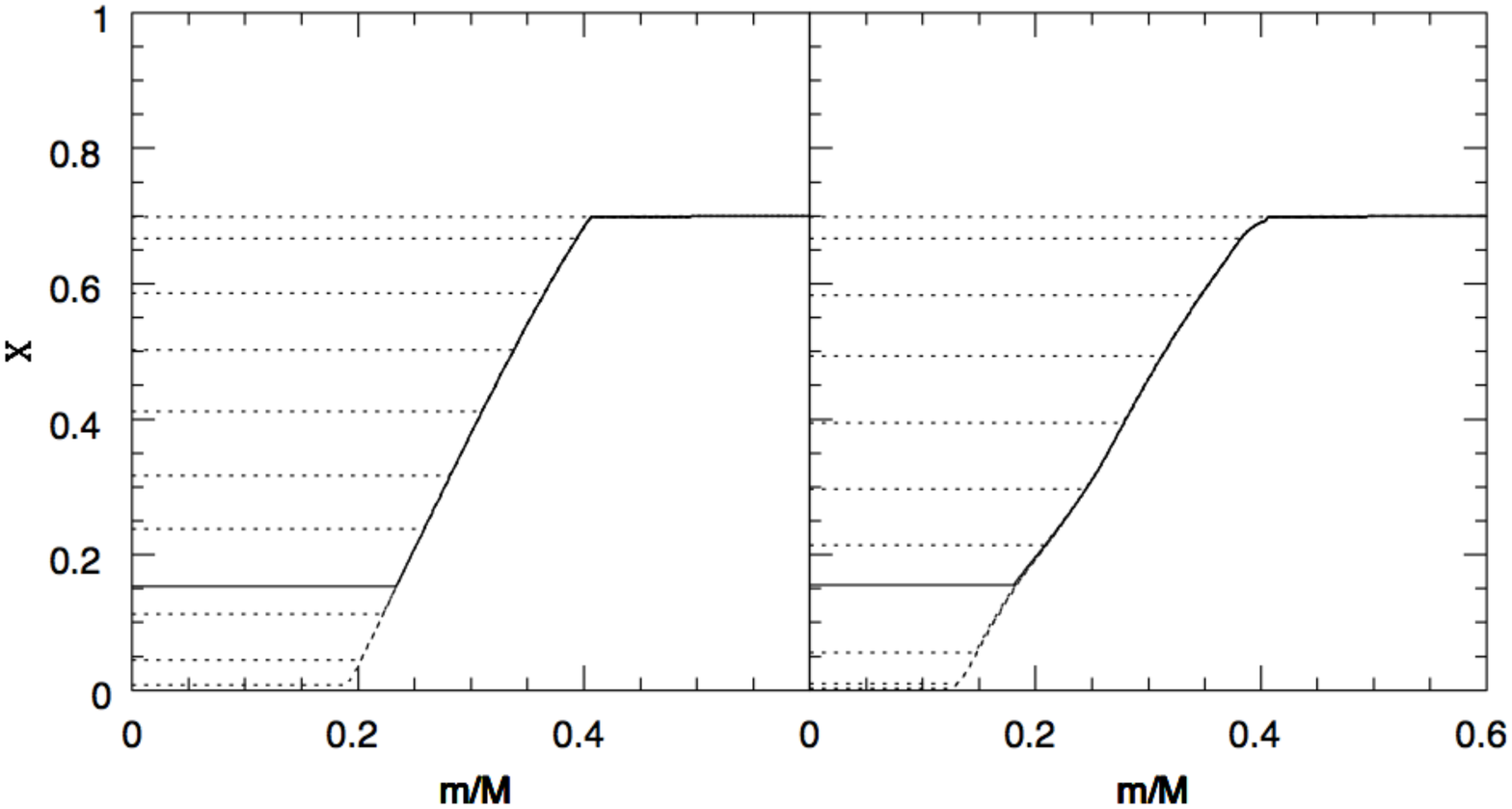} 
\caption{Hydrogen profile as a function of time in a 16 $M_{\odot}$ with a convective core boundary correctly computed (left panel) and computed with the method discussed in Sect. \ref{Misuse2} (right panel).The model displayed in Fig. \ref{bv_16M} is drawn in full line}
\label{X_16M}
\end{figure}

\begin{figure}
\hspace*{-4mm}
\includegraphics[scale=0.34,angle=270]{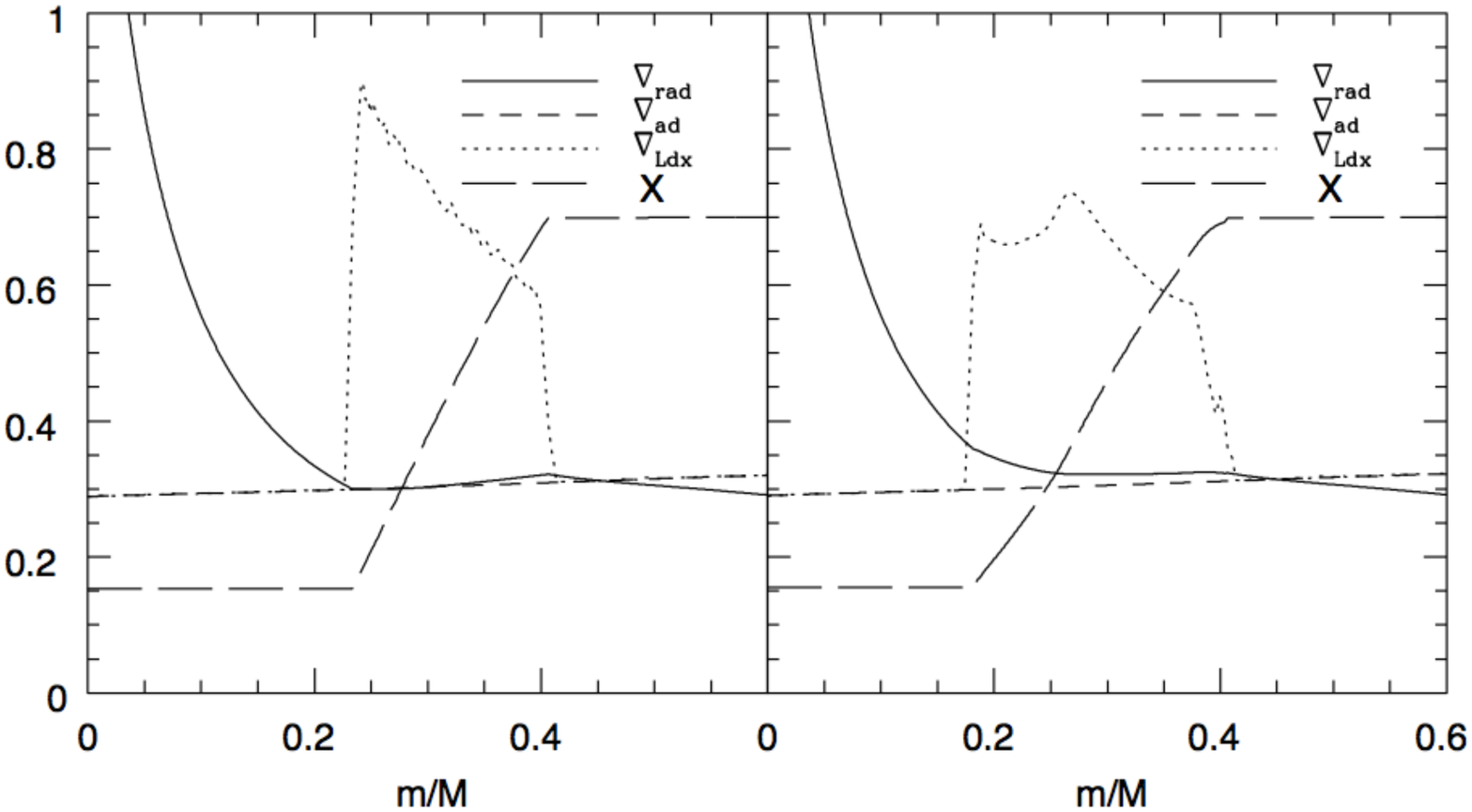} 
\caption{Hydrogen profile (long dashed line), radiative (full line), adiabatic (dashed line) and Ledoux (dotted line) temperature gradients for the MS model of 16 $M_{\odot}$ displayed in full line in Fig. \ref{X_16M}, correctly computed (left panel) and computed with the method discussed in Sect. \ref{Misuse2} (right panel), \rm using in \rm both cases the Ledoux criterion} 
\label{bv_16M}
\end{figure}

Such a case is illustrated in Fig. \ref{X_16M} where the time evolution of the hydrogen profile is shown as a function of the fractional mass for a 16 $M_{\odot}$ computed with the Ledoux criterion. In the left panel the correct solution, obtained with an extrapolation of $y$ from points in the convective core, is displayed while in the right panel the procedure discussed in Sect. \ref{Overview} and  \ref{Analysis} has been applied to fix the change of sign of $y$. One can see that, when reaching the end of core hydrogen burning, this star has a convective core too small by about 25\% (right panel) with the method discussed in Sect. \ref{Misuse2}. Fig. \ref{bv_16M} shows the hydrogen profile (long dashed line), the radiative (full line), adiabatic (dashed line) and Ledoux (dotted line) temperature gradients for the model drawn in full line in Fig. \ref{X_16M}. In the deep internal layers of the models discussed in this section, the actual temperature gradient is the adiabatic one in convective zones and the radiative one elsewhere. The location of the convective core boundary (at the limit of the mixed region) is indeed obviously incorrect in the right panel since $\nabla_R > \nabla_{ad}$ on the convective side of the boundary, which means $L_R < L(r)$ and $L_C > 0$.

Also, $y$ must vary very quickly in the first radiative shell above the core in order for it to cancel on the radiative side of its surface. This is physically meaningless since this implies that there $\frac{d\ln \mu}{dm} = 0$ and $\frac{dX}{dm} = 0$. But the derivative of X in the radiative zone is given by
\begin{equation}
\frac{dX}{dm} = \left \{ \frac{dX_C}{dt} \left [\frac{dm_C}{dt} \right ]^{-1} \right \}_{t_C} + \frac{d}{dm} \int_{t_C}^t \frac{dX(m)}{dt} dt
\label{dXdm}
\end{equation}
where $X_C$ and $m_C$ are respectively the hydrogen abundance in the core and its mass, while $t_C(m)$ gives the time when mass $m$ stops belonging to the convective core. Obviously $\frac{dX_C}{dt} \neq 0$ and $\frac{dm_C}{dt} \neq \infty$ and consequently $\frac{d\ln \mu}{d\ln P}$ is discontinuous there. Note that in that situation, the discontinuity of $\nabla_R - \nabla_{Ldx}$ increases very quickly since it is related to $(\frac{d\ln \mu}{d\ln P})_e = \frac{d\ln \mu}{d\ln m_C} (\frac{d\ln m_C}{d\ln P})_e$ and $\frac{d\ln \mu}{d \ln m_C}$ is large from the very beginning.

We also notice that in both models $\nabla_{ad} < \nabla_R < \nabla_{Ldx}$ in regions with a $\mu$-gradient. Such layers are semi-convective and they are more extended in the incorrect model since $\nabla_R - \nabla_{ad}$ is much larger. This is a direct consequence of the violation of the condition $\nabla_R = \nabla_{ad}$ (or $L_R = L$) on the convective side of the core boundary. In the correct model and within its accuracy, $\nabla_R = \nabla_{ad}$ in the layers above the convective core (however their extent decreases as the mass increases and they have nearly disappeared in a 40 $M_{\odot}$ star with the same central hydrogen abundance) and $\nabla_R > \nabla_{ad}$ is found only in layers close to the homogeneous envelope.

Also a small shell with $\nabla_R > \nabla_{ad}$ is found in the homogeneous region just on top of the ZAMS core and it should be handled as a convective shell. Its extent also grows with the mass of the star. This would however raise problems as its bottom is not convectively neutral. It was shown by \citet{Gabriel70} that the overshooting at its bottom leads to mixing and to a downwards displacement of the shell able to form a semi-convective region with a chemical composition such that the Schwarzschild criterion is fulfilled everywhere. However, the end result is very sensitive to the ability of overshooting to produce mixing and Gabriel's estimate was very likely overoptimistic. Later on, he made a less optimistic estimate and reached the same conclusion \citep{Gabriel95}. However, only numerical simulations of such situations seem able to provide a reliable answer. This has so far not yet been done.

\rm But we want to call attention on a point which is nearly always ignored~: it is obvious that the occurrence of semi-convective regions in stellar models is directly related to the dependence of the opacity not only on the pressure and the temperature but also on the chemical composition. Therefore this latter dependence must be taken into account in numerical simulations and in theoretical studies to be relevant for stellar models. \rm 

Fig. \ref{bv_16M} also shows that $\nabla_{Ldx}$ is very different in the two panels. This means that the Brunt-V\"ais\"al\"a frequencies in the two models will significantly differ with all the consequences this may have for non radial oscillations studies.

\subsubsection{The convective core grows during MS} \label{ccgrow}

\rm The problem is still worse when the convective core mass increases with time because most codes use the core mass of the last computed model as starting guess for the new one. As a result the algorithm will always provide too small a trial value of the core mass and it will quickly be unable to increase it up to its correct value. As soon as the discontinuity of $y$ becomes large enough to meet case 3 (Fig. \ref{graph12}, left panel) at the first iteration, the boundary will stick to the initially guess, \rm making impossible to obtain \rm any further significant increase of the core mass. \rm 

At the beginning of the MS phases and whatever the stellar mass, the convective core mass increases up to its maximal extent. In stars massive enough however this maximum is reached very quickly before a significant change in $\mu$ (and so a discontinuity of $y$) can build up and the maximum extent of the core is very close to the correct one (see Fig. \ref{X_16M}).  

In low mass stars with a small convective core the situation is very different from that encountered in more massive ones since the convective core mass increases much more slowly. At the same time a $\mu$-discontinuity builds up at the base of the radiative layers and a $\mu$-gradient, resulting from nuclear reactions outside the convective core, slowly forms. Moreover when the Ledoux criterion is used, with the procedure discussed in Sect. \ref{Misuse2}, to fix the location of the convective core boundary, $\nabla_{Ldx} - \nabla_{ad}$ soon becomes large enough to meet case 3 (Fig. \ref{graph12}). This prevents any later growth of the convective core mass long before it reaches its correct maximum value. Later on it goes on shrinking and rapidly vanishes. 

\begin{figure}
\hspace*{-6mm}
\includegraphics[scale=0.34,angle=270]{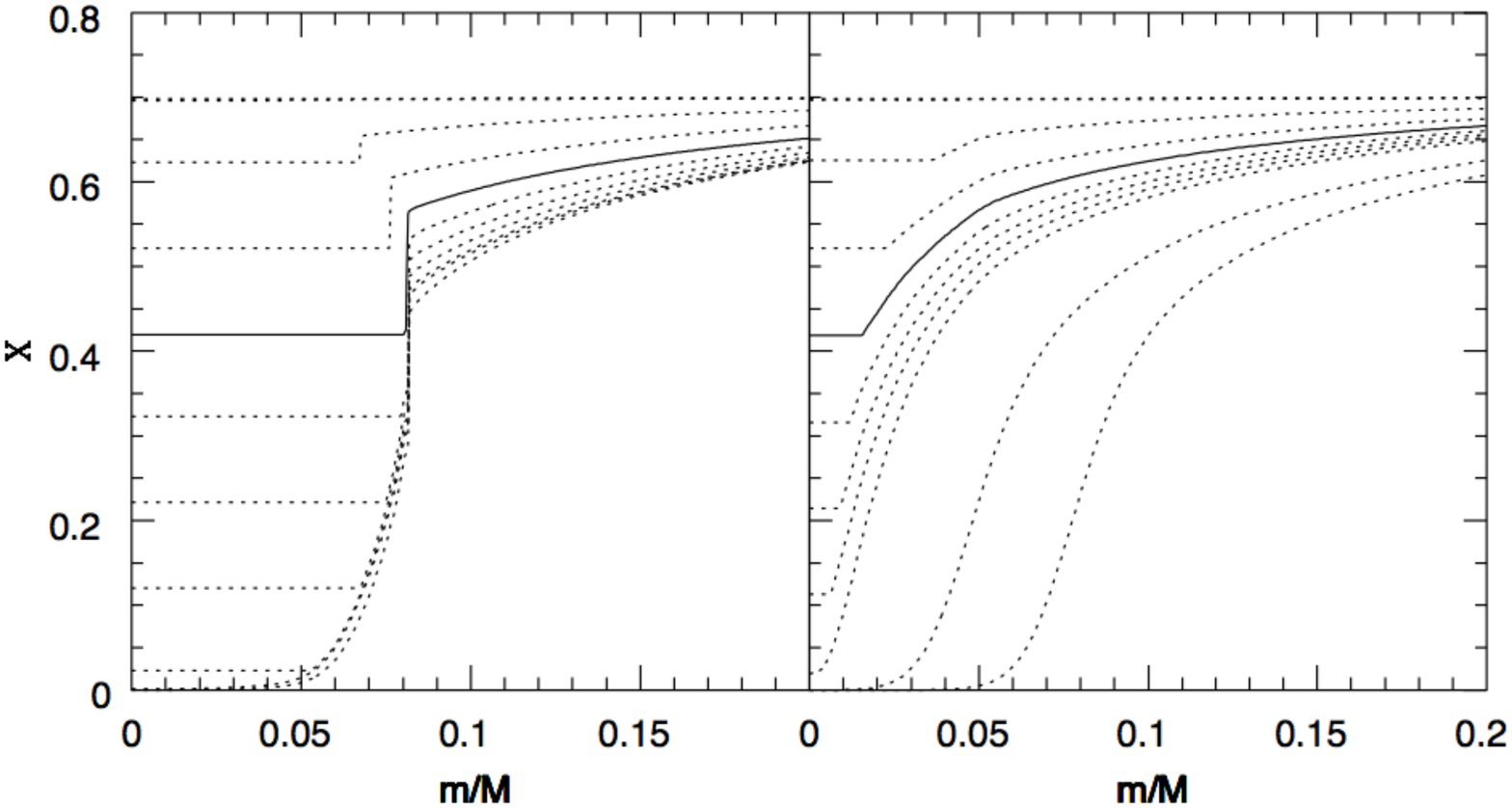} 
\caption{Hydrogen profile as a function of time in a 1.5 $M_{\odot}$ correctly computed (left panel) and computed with the method discussed in Sect. \ref{Misuse2} (right panel). The model displayed in Fig. \ref{bv_1.5M} is drawn in full line}
\label{X_1.5M}
\end{figure}

\begin{figure}
\hspace*{-4mm}
\includegraphics[scale=0.33,angle=270]{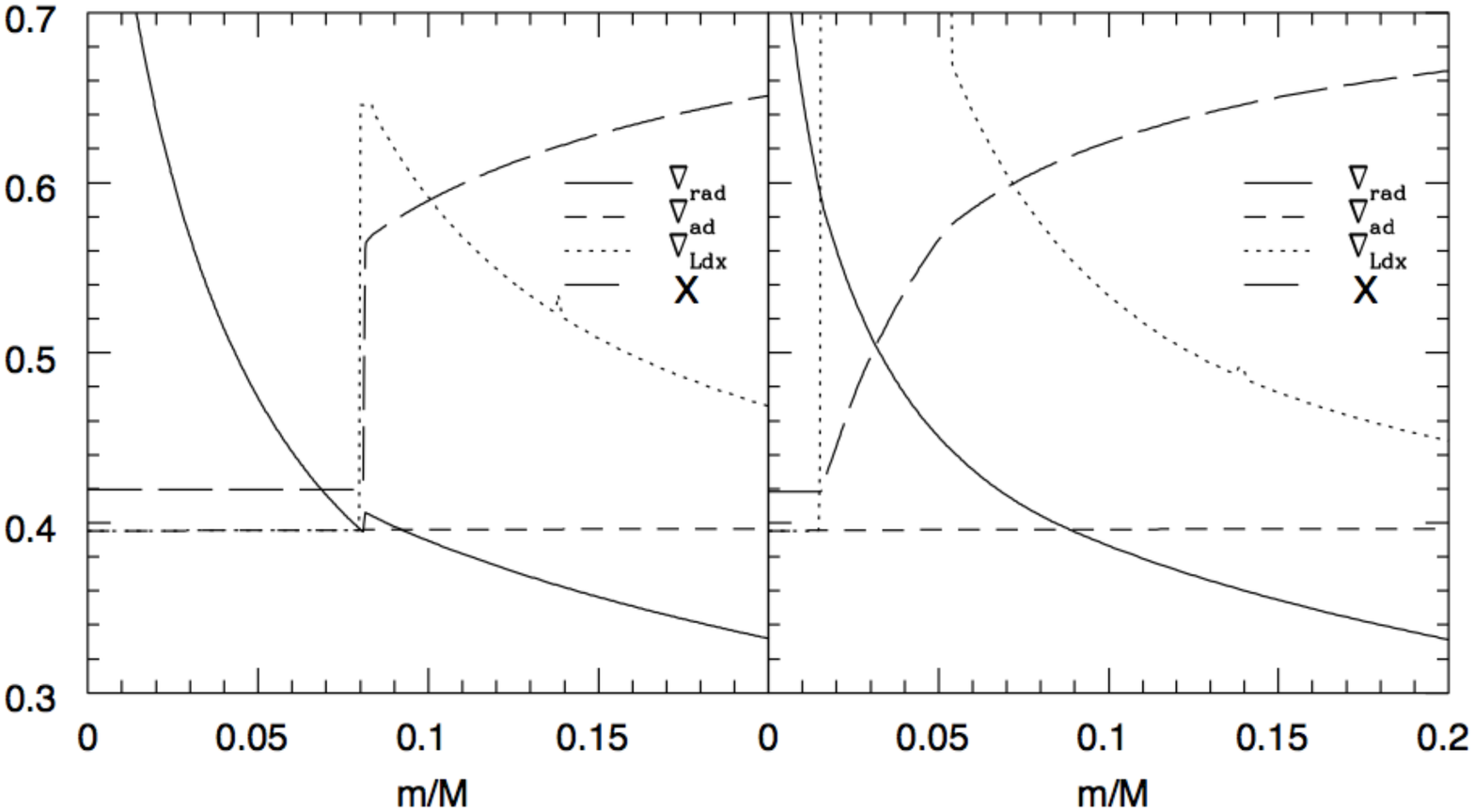} 
\caption{Hydrogen profile (long dashed line), radiative (full line), adiabatic (dashed line) and Ledoux (dotted line) temperature gradients for the MS model of 1.5 $M_{\odot}$ displayed in full line in Fig. \ref{X_1.5M}, correctly computed (left panel) and computed with the method discussed in Sect. \ref{Misuse2} (right panel), using in both cases the Ledoux criterion} 
\label{bv_1.5M}
\end{figure}

This situation is illustrated in Figs. \ref{X_1.5M} and \ref{bv_1.5M} for a 1.5 $M_{\odot}$ correctly computed through extrapolation from the convective core (left panels) and computed with the procedure discussed in Sect. \ref{Misuse2} (right panels). Both evolutionary sequences were computed with the Ledoux criterion. 
The extent of the convective core (shown by the extent of the mixed region) is very different; the core even rapidly becomes very small when its boundary is incorrectly determined \citep[see also for example the results obtained by][Fig. 13]{Silva11,Paxton13}. It may also remain stuck at a maximum value reached before a significant $\mu$-discontinuity has time to develop, though with $\nabla_R > \nabla_{ad}$ at the convective boundary (as in \rm the CESAM code, \rm Y. Lebreton, private communication). 

From Fig. \ref{bv_1.5M} showing the hydrogen profile (long dashed line), the radiative (full line), adiabatic (dashed line) and Ledoux (dotted line) temperature gradients for the model drawn in full line in Fig. \ref{X_1.5M}, it is obvious that the convective core boundary is incorrect in the right panel since $\nabla_R$ is significantly larger than $\nabla_{ad}$ at the boundary. The same problem shows up in \citet[][Fig.1]{Silva11}. As in more massive stars, discussed above with relation \ref{dXdm}, the Ledoux gradient is very different in the radiative shell just above the core.

Note that in these stars the opacity is larger just outside the core than just inside it because of the discontinuity of chemical composition. This does not affect our analysis since these models were computed with the Ledoux criterion and the stabilizing influence of the $\mu$-gradient is much greater than the destabilizing one caused by the opacity increase. However $\nabla_{ad} < \nabla_R < \nabla_{Ldx}$ in some layers just above the convective core as can be seen in the left panel of Fig. \ref{bv_1.5M} (they correspond to the layers between points $A$ and $B$ in the right panel of Fig. \ref{scheme}).
Again this region is more extended in the right panel because in that model $\nabla_R$ is much larger than $\nabla_{ad}$ at the surface of the convective core.
They are often considered as semi-convective and they disappear in the correct models when the convective core starts shrinking. 

The same problem is encountered at the bottom of a red giant convective envelope as soon as a $\mu$-discontinuity is present at the boundary; an incorrect use of the Ledoux criterion will quickly stop any further downwards extension of the envelope.

In low mass stars computed with the Schwarzschild criterion, we are in the situation considered 
on the right panel in Fig. \ref{scheme}. This is the only situation we know of where the Schwarzschild criterion leads to a positive discontinuity of $y$. With the correct method the core boundary is located at point $A$. The discontinuity of $y$ is there and also at that point the solution jumps from the lower curve to the upper one. As a result the model seems inconsistent since $y > 0$ in a non negligible region above point $A$ extending up to point $B$. However the Schwarzschild criterion
is meaningless to check the stability there since  a $\mu$-gradient is present in that radiative region and is large enough to give $\nabla_{ad} < \nabla_R < \nabla_{Ldx}$.

If the core boundary is placed at the point where $y=0$ in the radiative envelope, then it will be found at point $B$. The discontinuity of $y$ is there and again, since it is positive, the solution jumps from the lower curve to the upper one when moving outwards. If the consistency of that model is checked it will be found inconsistent since $y<0$ in the upper layers of the convective core and, since the core is chemically homogeneous, there is really a problem with such a model. This is the method followed in the ASTEC code \citep{Christensen08}. The layers between point $A$ and $B$, although convectively stable are mixed with the convective core.

Let us now see what happens when the method discussed in Sect. \ref{Misuse2} is used. First let us assume that the initial core mass is larger than $m_B$  (right panel in Fig. \ref{scheme} - case 2). After the first iteration $y$ changes sign in the convective core (close to point $A$) where it is continuous. The core boundary is now put close to its correct value for the second iteration. However an interpolation never gives the perfect solution and more important during stellar model computations we are dealing with approximate solutions of the equations and not with relaxed models. As a result, the mass of the convective core obtained for the second iteration will turn out to be either too small or too large at the end of that iteration. If it is too large, there is at least one point in the core with $y < 0$ and we
switch to case 4 (see below). It is nevertheless possible to find a new core mass close to point $A$ for the next iteration and there is a chance that it leads to a relaxed model. However that model has a convective shell. 
On the contrary, if it is too small, then $y > 0$ at the surface of the core, we are now in case 3 discussed below. As a result, at the third iteration, the core boundary will be moved close to point $B$. 

In case 3, the initial core mass is smaller than $m_A$. Then after the first iteration $y$ changes sign close to point $B$ and much too large a core mass is used for the second iteration. But again since convergence will generally not be obtained at the end of the second iteration, what follows depends on the sign of $y$ at the surface of the convective core. If $y>0$ at the end of that iteration, we switch to case 4. It is nevertheless possible to find a new core mass close to point $B$ for the next iteration and there is a chance that it leads to a converged model. But if $y<0$  we are back to case 2 discussed above and at the next iteration the core boundary will be moved close to point $A$.
We see that in both cases 2 and 3 the core boundary may oscillate  between points $A$  and $B$ until, by chance, a converged model is found. Unfortunately the final core mass could be close to either point $A$ or $B$.

In case 4 the assumed core boundary is between points $A$ and $B$. $y$ changes sign three times in the initial guessed model. 
In the first iteration step, the model has a convective core which has about the right mass (close to point $A$), then a radiative shell followed by a convective one and finally a radiative envelope. What follows after the end of that iteration depends again on the sign of $y$ at the last point in the convective core. If $y < 0$, we remain in case 4 with about the same core mass and the model has a chance to converge (but it has an extended convective shell). If $y > 0$, we are back to case 3 and the core
boundary will be moved close to point $B$ at the next iteration (see above). Again the core mass will oscillate between points $A$ and $B$ until by chance (or because of some trick used in the code) a converged model is found. 

However some codes choose to ignore convective shells. Then in case 4, what happens depends on the way the code searches for the zero of $y$, either starting from the center outwards or starting from the surface inwards. 
In the first hypothesis a core boundary close to point $A $ will be found for the first iteration; with the second one it will be close to point $B$. With both methods, what happens next depends upon the sign of $y$ at the core boundary after the next iteration and the same kind of discussion as above can be made. The final result will again be a model with a core boundary found by chance close to either point $A$ or point $B$ \citep[][Fig. 7]{Montalban07}.

The method used in ASTEC \citep[see above][]{Christensen08}, though leading to inconsistent models, prevents such oscillations of the convective core boundary. It leads to a convective core mass equal to the external envelope of the oscillating core mass just here above discussed \citep[see the ASTEC curve in Fig. 9 in][]{Lebreton08}.   

\subsection{Core helium burning}
The convective core also expands during a large fraction of core He burning leading to a $\mu$ discontinuity but to no $\mu$-gradient. Again if the same core mass as in the previous model is taken as the initial guess, the algorithm will quickly predict too small a core mass. As the discontinuity of $y$ increases progressively with time, it will become large enough to meet case 3 at the first iteration. It will then be impossible to significantly increase the core mass any further \citep[see for example Fig. 15 in][]{Paxton13}. 

\begin{figure}
\hspace*{-4mm}
\includegraphics[scale=0.33,angle=270]{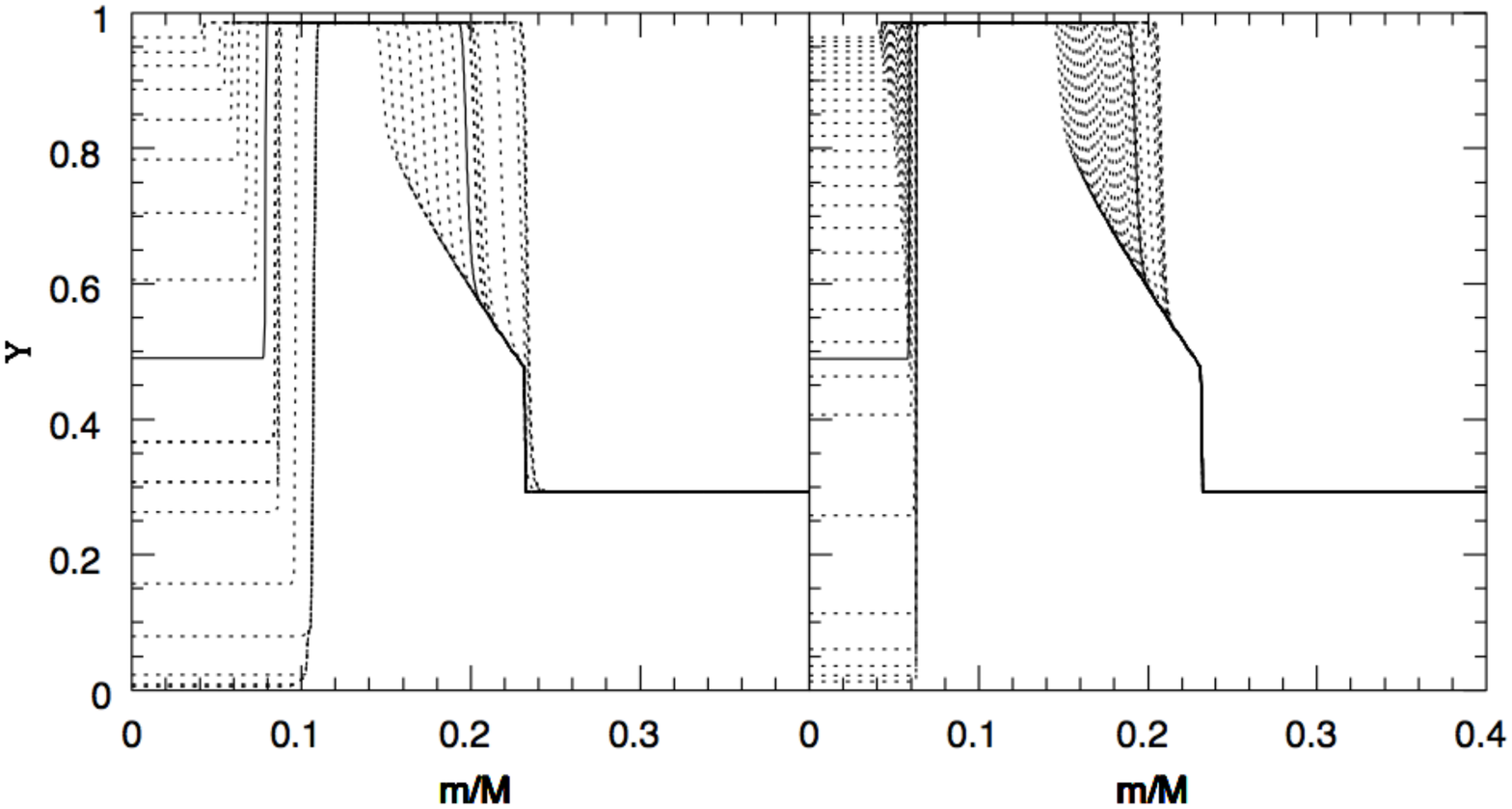} 
\caption{Helium profile as a function of time in a core helium burning 8 $M_{\odot}$ star correctly computed (left panel) and computed with the method discussed in Sect. \ref{Misuse2} (right panel). The model displayed in Fig. \ref{bv_8M} is drawn in full line} 
\label{Y_8M}
\end{figure}

\begin{figure}
\hspace*{-4mm}
\includegraphics[scale=0.33,angle=270]{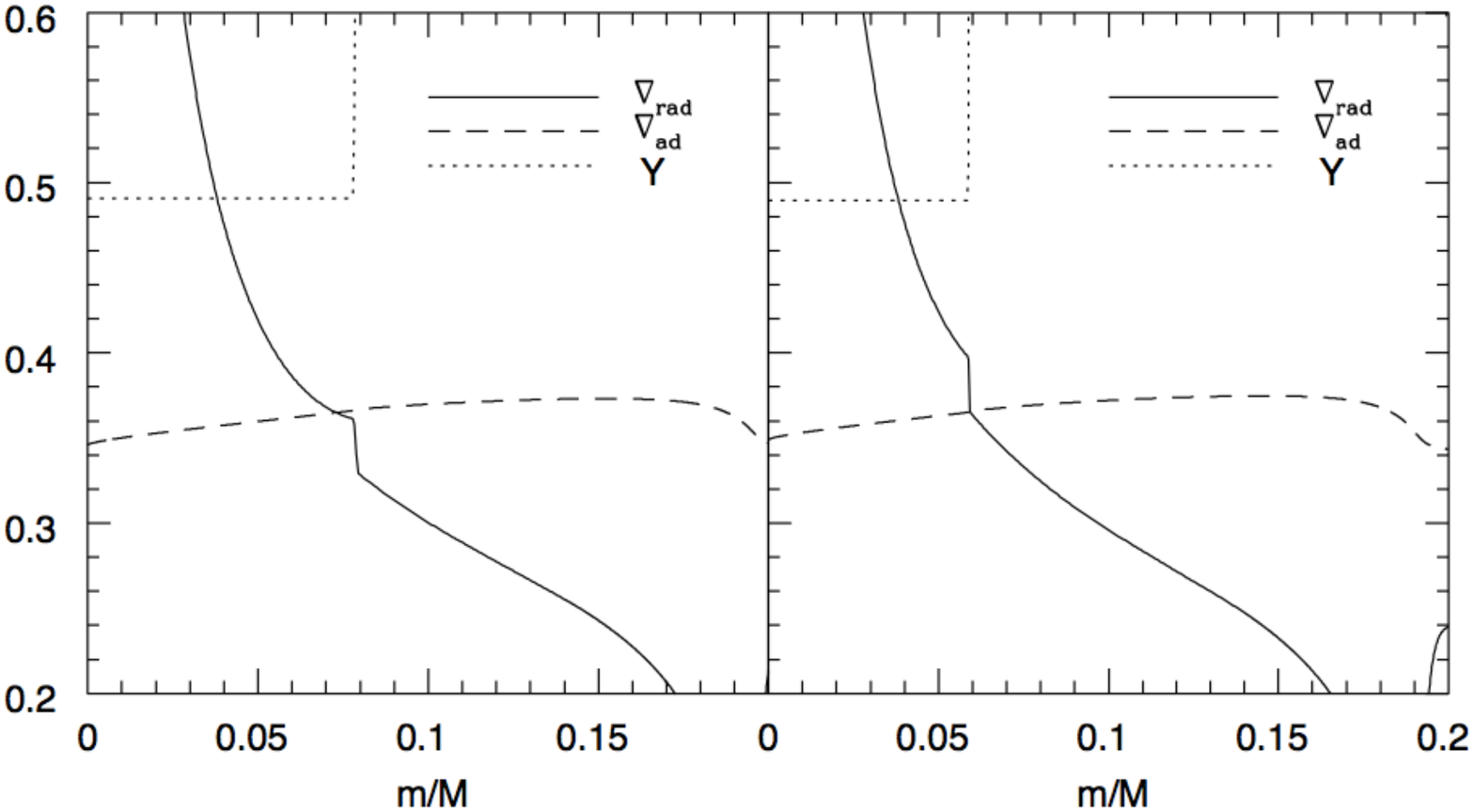} 
\caption{Helium profile (dotted line), radiative (full line) and adiabatic (dashed line) temperature gradients for the core He-burning model of 8 $M_{\odot}$ displayed in full line in Fig. \ref{Y_8M}, correctly computed (left panel) and computed with the method discussed in Sect. \ref{Misuse2} (right panel)} 
\label{bv_8M}
\end{figure}

This is illustrated in Fig. \ref{Y_8M} showing the helium profile as a function of time in a 8 $M_{\odot}$ core helium burning star computed with a correct location of the convective core (left panel) and with the method discussed in Sect. \ref{Misuse2} (right panel). The extent of the convective core is significantly larger when a correct positioning is adopted. Fig. \ref{bv_8M} shows the mass distribution of helium, and of the radiative and adiabatic temperature gradients in the model drawn in full line in Fig. \ref{Y_8M}. One can clearly see in the right panel, that $y > 0$, or $\nabla_R > \nabla_{ad}$ at the convective side of the core boundary, which means $L_R < L(r)$ and $L_C > 0$. This is the problem noticed by \citet{Castellani71b}.

To summarize, the algorithm discussed here above \rm may in some situations lead \rm to solutions close to the right one but in many others to completely incorrect ones.

\section{Double mesh point at convective boundaries} \label{Double} 
The best way to deal more easily with the special properties of a convective zone boundary is to systematically introduce a double mesh point at each boundary. A discontinuity in chemical composition always leads to a discontinuity in density. If only one mesh point is present at such a discontinuity, the chemical composition and the density may not be clearly specified. Two mesh points, each with the same mass but with the chemical composition and density of each sides of the discontinuity removes the ambiguity. Since the discontinuous variables appear in different places of the system of differential equations, their values must be properly known in order to correctly write the set of difference equations in the layers just above and just below the discontinuity. This is easily done when a double mesh point is introduced at the discontinuity. 
 
A single mesh point or even an enlarged density of mesh points over some interval is unable to correctly stand for a discontinuity. \rm For instance, the discontinuity in density is then replaced by a very steep $\mu$-gradient in the shell adjacent to the convective zone. This leads in that shell to a stable stratification according to the Ledoux criterion while, when the discontinuity is properly taken care of, that term is much smaller. \rm For instance during the central helium burning phases, there is practically no $\mu$-gradient above the expanding convection core as it grows in practically pure helium layers. This is what is correctly found when a double mesh point is introduced at the surface of the core since there is no significant $\mu$-gradient term in the Ledoux criterion above the core while without a double mesh point, there is a large $\mu$-gradient in just the one shell which is on top of the convection core. As a result, the Ledoux criterion predicts a much stronger stability just in that shell with the consequence that the convection core quickly starts shrinking to finally disappear. This explains the difference in core mass found by \citet{Paxton13} in their Fig. 15 when the Schwarzschild or the Ledoux criterion is used during core helium burning.

Moreover, as stars evolve, the displacement of these boundaries may in some cases leave behind a discontinuity in a radiative zone. This occurs for instance during the red giant phase after the first dredge-up. When the discontinuity is located in a radiative zone, a double mesh point is even more important, not only for the reason given above but mainly to avoid destroying the discontinuity. Without a double mesh point the discontinuity is progressively smoothed out by the addition and retrieval of mesh points in a sort of numerical diffusion. This may completely change the chemical composition profile in an undue way indeed.

Even when the chemical composition is continuous at the boundary of a convective zone, a double mesh point is useful when there is a $\mu$-gradient in the adjacent radiative layers since this $\mu$-gradient is discontinuous at the convective zone boundary and therefore also the density gradient as well as the Brunt-V\"{a}is\"{a}l\"{a} frequency which is used for non-radial oscillations studies.

To our knowledge, the code developed by L. Henyey and his group is the only one to have introduced double mesh points in all the situations considered in this section. When there is a discontinuity in the chemical composition, the CESAM code developed by P. Morel also introduces a double mesh point at the boundary of a convective zone.

\section{Problems raised when a convective shell sets in} \label{Shell}
Let us assume that, in some part of a region with a gradient in chemical composition, supposedly in radiative equilibrium, the Ledoux criterion predicts instability.  Those layers become convective and should be mixed. The problem is now to find the position of the boundaries which both verify the condition $\nabla_R = \nabla_{ad}$  on the convective side. The end result will be a convective shell which has different boundaries than those initially found. This may not be an easy task for the code to find a solution, on the assumption that it exists.  But assuming that such a solution does exist, it is still necessary to check its consistency. 

\begin{figure}
\includegraphics[scale=0.25,angle=270]{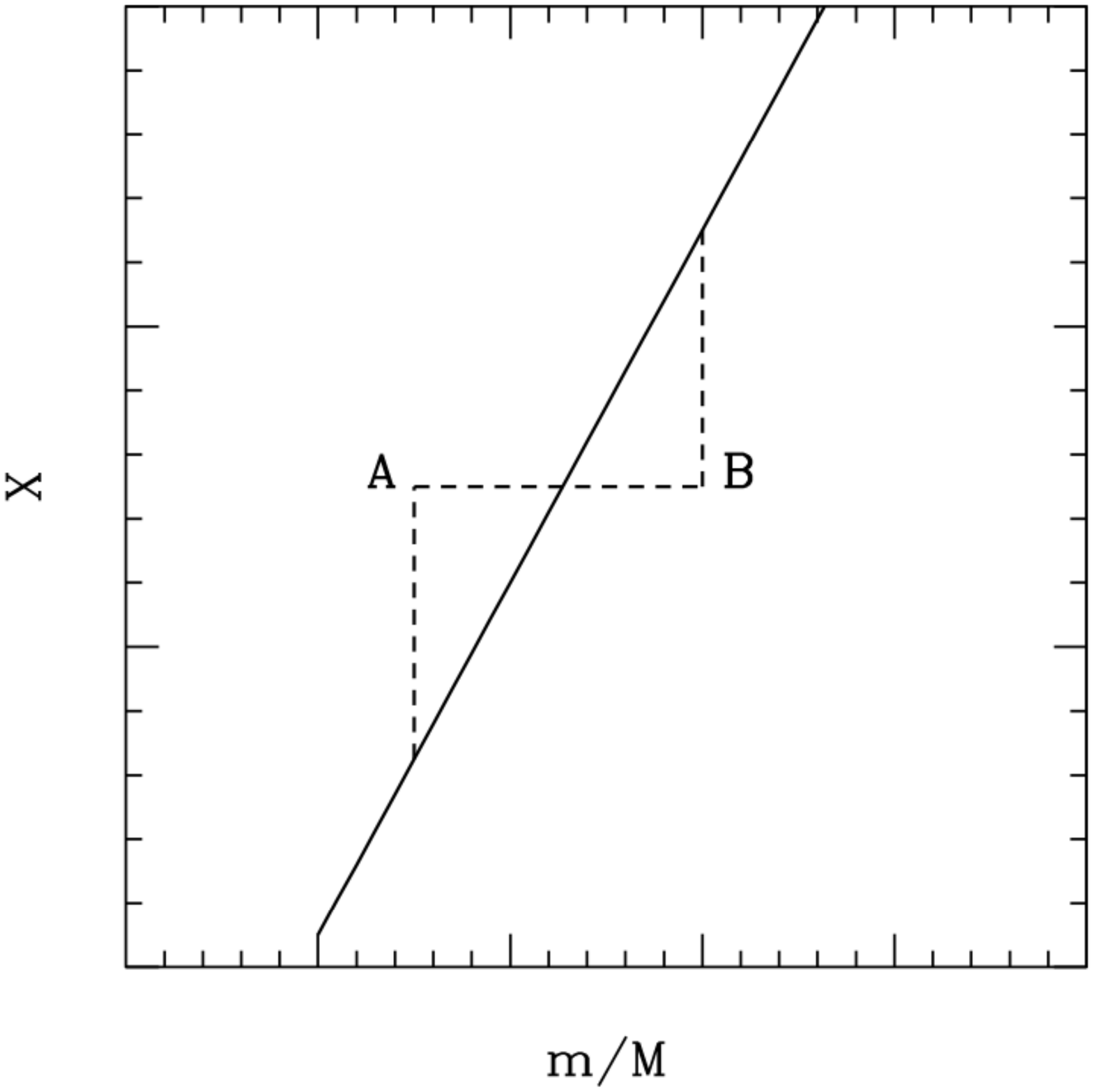} 
\caption{Modification of the $X$ profile when a convective shell $AB$ sets in}
\label{graph4}
\end{figure}

At points $A$ and $B$ in Fig. \ref{graph4}, on the radiative side of the discontinuity, we have $\nabla_R =  \frac{\kappa_e}{\kappa_i} \nabla_{ad}$. In most cases  $\frac{\kappa_e}{\kappa_i} > 1$  at one of the two boundaries.  For a coherent model to be found, the Ledoux criterion must predict stability on the radiative sides of the discontinuities. Therefore at one of the two boundaries the temperature gradient is in between the Schwarzschild and Ledoux critical values. It is well known that a semi-convection zone may develop in such a situation. It will not necessarily be so but the question needs to be studied each time it is encountered.

\begin{figure}
\includegraphics[scale=0.4,angle=0]{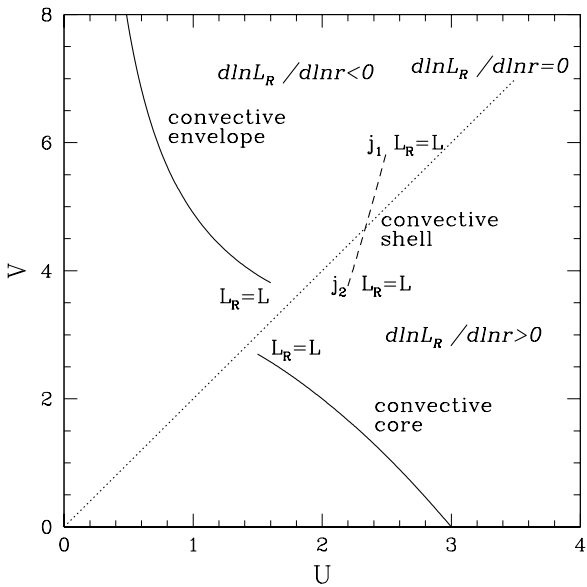} 
\caption{Schematic solutions for a convective core, a convective envelope and a convective shell in the (U,V) plane}
\label{graph5}
\end{figure}

But worse, it is not obvious that a static solution may exist. In other words, it might be impossible to fulfill the condition $\nabla_R = \nabla_{ad}$ at both boundaries of the convective shell. Inside a convection zone we have
\begin{equation}
L_R = \frac{16 \pi a c G T^4}{3 \kappa P} m \nabla_{ad} \;\;.
\end{equation}
Taking the derivative, we end with
\begin{equation}
%\begin{array}{lll}
\frac{d\ln \frac{L_R}{L}}{d\ln r} = U + \left [1 + \frac{\partial \ln \frac{\kappa}{\nabla_{ad}}}{\partial \ln P} + \left ( \frac{\partial \ln \frac{\kappa}{\nabla_{ad}}}{\partial \ln T} -4 \right ) \nabla_{ad} \right ] V \\
- \frac{d \ln L}{d \ln r} \label{UV}
%\end{array}
\end{equation}
where U stands for ($\frac{d \ln m}{d \ln r}$) and V for  ($- \frac{d \ln P}{\ln r}$). The curve along which $\frac{d \ln L_R}{d \ln r}$  is equal to 0, assuming that the total luminosity is constant, is illustrated by the dotted curve in Fig. \ref{graph5}. In  most cases the derivatives of the opacity as well as of the adiabatic gradient and the luminosity will change only slightly over the extent of the shell.

It is easily seen that $\frac{d \ln L_R}{d \ln r}$ is positive below that curve and of course negative above it.

Let us consider the very often encountered case of a convective core large enough so that there is no nuclear reactions in the vicinity of its surface. It is necessary that its surface be reached when the solution is still below the critical line since if $L_R < L$ when its maximum value is reached, the condition $L_R = L$ will never be fulfilled at its surface. 

In a convective envelope while progressing inwards, $L_R$ increases as long as the solution is above the critical line since there $\frac{d \ln L_R}{d \ln r} < 0$ and its bottom must be reached before the envelope solution crosses that critical line.

As far as a convective shell without nuclear reactions is concerned, its inner boundary must be above the critical line and its outer one below it, in order that the condition $L_R = L$ be fulfilled at both boundaries. At the bottom ($j_1$) $L_R = L$, then $L_R$ decreases until the critical curve is reached and afterwards it starts increasing to eventually reach $L_R = L$ at the top ($j_2$). This requires a very special position of the shell and there is \textit{a priori} no reason that any shell can satisfy this condition. In most cases $L_R$ will indeed be a monotonic function of the position within a convective shell. 

There are however two cases where boundaries for a convective shell should be found without any problem. The first one is when the convection zone is caused by an opacity bump, related for instance to the ionization of a chemical element. The derivatives of the opacity drastically change throughout that region and it is possible to have $\frac{d \ln L_R}{d \ln r} < 0$ at the bottom and of the opposite sign at the top. The second one is when the convection zone appears within a shell where nuclear reactions significantly increase the luminosity. Then the term $\frac{d \ln L}{d \ln r}$ that we have neglected in our discussion becomes important and strongly varies with the position in the shell.

What happens when a static solution cannot be found? The convective velocity is then positive at one boundary of the shell and overshooting must take place, bringing in material with a different chemical composition. The result is a change in the chemical composition of the shell, which will move. This might be a way to produce a semi-convective zone (this is the same mechanism as that proposed by \citet{Gabriel70} for massive main sequence stars). However the difficult and yet unsolved question is to compute the speed of this shell motion.

\section{A few remarks concerning overshooting} \label{Over}
First it must be realized that there are two kinds of overshooting problems. The first one is encountered when convective motions penetrate stable layers with the same chemical composition. In this situation, convective material is progressively slowed down and is eventually thermalized. It has then the same density as the surrounding material so that it has no natural tendency to move backwards. It also applies to a shrinking convective core provided that the mass on top of the overshooting zone also decreases and does not itself become convectively unstable. \rm These problems have somewhat been discussed theoretically but the number of numerical simulations and of situations studied with the moment theory are still too few to allow finding a rule to specify the depth of the overshooting (undershooting) layers and the run of the temperature gradient (see for instance \citet{Kupka02, Kupka09, Montgomery04, Zhang12b, Zhang12a, Zhang13} and, for numerical simulations, \citet{Freytag96, Freytag04, Kochukhov07, Rogers06, Tian09, Viallet13}). \rm 

When it is possible to use better theories, the usual boundary condition (\ref{Sch}) becomes meaningless and the more fundamental one \ref{cond}, or its equivalent $V_r = 0$, has to be used. Of course they do not apply outside convective zones. But presently, we can only introduce an overshooting (undershooting) region of arbitrary thickness, $d_o$, above (below) the point where $\nabla_R = \nabla_{ad}$. This point is now located in a chemically homogeneous region. It follows that $\nabla_{Ldx} = \nabla_{ad}$ there and that the function $y= \nabla_R - \nabla_{ad}$ has of course no discontinuity. As a result methods considered in Sects. \ref{Def} and \ref{Misuse2} give the same result and there is no problem anymore. However, we have to realize that $\lim_{d_o \rightarrow 0} M_C(X_C,d_o)$ is equal to the function $M_C(X_C)$
obtained with the exact method of Sect. \ref{Def} and not with that given by Sect. \ref{Misuse1} and \ref{Misuse2}. As a result, when it comes to discussing the influence of overshooting on some observables, Sect. \ref{Misuse1} and \ref{Misuse2} offers a wrong zero point.

\rm The second kind of overshooting problem occurs when material overshoots in layers with a different chemical composition. This occurs for instance when the core expands or also during the first dredge-up if undershooting occurs at the bottom of the convective envelope. Let us just discuss the problem for a convective core. Then in the overshooting layers, the raising material is immediately significantly heavier than its surrounding and it is quickly slowed down by buoyancy forces. Moreover, even if it advects some surrounding material, its molecular weight remains larger than that of the surrounding. Therefore, when it stops moving upwards, it comes naturally back downwards even if it succeeds to reach the same temperature as the surrounding material. We therefore expect a much smaller overshooting than when there is neither a gradient nor a discontinuity in molecular weight. Indeed \citet{Canuto98, Canuto99b} demonstrated that  the extent of overshooting is smaller when it takes place in a region with a gradient in molecular weight. However the author does not specify the amount of that decrease. These who have to compute stellar models have thus two possible choices~: either they solve the system of equations given by Canuto (unfortunately nobody does that), or they guess the value of the overshooting extent (the way out always adopted). \rm 

\section{Conclusions} \label{Conclusions}
Following a growing uneasiness among stellar modelers as to what should be the correct implementation of a reliable algorithm to fix the boundaries of convection zones in stars, we have attempted to discuss and clarify some related important points. First we have recalled the physical aspects of a convective zone and its boundaries and we have given the only correct way to find them. We have then shown that there are two possible ways of misusing the convective zone boundary condition and we have discussed the consequences for models computation. Our main conclusions are the following:
\begin{enumerate}
\item The neutrality condition $L_R = L$, or $\nabla_R = \nabla_{ad}$ when LMLT is used, must be applied \textit{on the convective side} of a boundary and, during the iteration process, the improvement of a boundary location must imperatively be done through extrapolations or interpolations from points \textit{in the convective zone only} in order to find the point for which $L_R = L$. This result obtained in Sect. \ref{Def} stands on a very firm basis and can hardly be put into question. It allows a very simple test to anyone computing stellar models. Just plot either $L - L_R$ or $\nabla_R - \nabla_{ad}$ throughout the models; if this does not cancel out at each boundary of convective zones (within the accuracy required for converged models), then there is a problem with the algorithm used to locate them. \\

\item The neutrality condition which can then be chosen either as $y = \nabla_R - \nabla_{Ldx} = 0$ or $y = \nabla_R - \nabla_{ad} = 0$ \textit{may never be applied on the radiative side} when the checked variable is discontinuous at the boundary. More precisely, during the iteration process the improvement of a boundary location \textit{may not} be done through extrapolations or interpolations from points in the radiative
zone in order to find the point for which either $\nabla_R - \nabla_{Ldx} = 0$ or $\nabla_R - \nabla_{ad} = 0$ (see Sect. \ref{Misuse1}). \\

\item When the convective boundary is searched for through a change of sign of $y$ ($y = \nabla_R - \nabla_{Ldx}$ or $y = \nabla_R -\nabla_{ad}$) (see Sect. \ref{Misuse2}), we have more explicitly discussed the situation for a convective core. Two cases have to be considered separately:

\begin{enumerate}
\item	 At the core boundary $y$ is larger on the convective side than on the radiative one of the discontinuity. Two possibilities arise at each step of the iteration process. If the estimated convective core is too small the algorithm will generally predict an incorrect boundary. If it is too large, its extent will be correct provided the estimated core mass remains always too large but it will be underestimated if during the iterations an oscillation in the core mass around the correct value occurs. \\

\item	At the core boundary $y$ is smaller on the convective side than on the radiative one of the discontinuity. Then the position of the core boundary may, just by chance, be either close to the correct value or much larger. \\

\end{enumerate}

\item The best way to avoid inconsistencies in the definition of variables appearing at a convective boundary is to add a \textit{double} mesh point at the exact location of the boundary with the neutrality condition satisfied on the convective side of the double mesh point. If there is a discontinuity in the chemical composition at the convection zone boundary, then the inner mesh point will have the same chemical composition as the convective zone and the upper one that of the outer radiative region (see Sect. \ref{Double}). \\

\item Since the location of convective boundaries must be found from extrapolations from \textit{inside} a convective zone (see points 1 and 2), the Ledoux criterion is only required to check the appearance of a convective layer in an otherwise radiative zone located in a $\mu$-gradient region. \\

\item If such a convective shell appears, it could be impossible to locate its boundaries in such a way that  the neutrality condition is satisfied on the convective side of each of them and then it is possible that a semi-convective layer forms just above (or below) its upper (lower) boundary (see Sect. \ref{Shell}). \\

\item The correct application of the issues solved in this paper is an \textit{essential pre-requisite} to any attempts of introducing more sophisticated physics, and to any attempts or claims at testing "stellar model/physics" using observations (seismology of course is presently the most promising, but not the only one). \\

\end{enumerate} 

\begin{acknowledgements}
The authors wish to especially thank Pierre Morel for long and interesting discussions which helped improving significantly several of the arguments presented in this paper. They are also grateful to Richard Scuflaire and Marc-Antoine Dupret for their careful rereading of the manuscript and for their pertinent remarks. Many thanks to Yveline Lebreton who computed for us some MS low mass star models with CESAM.
\end{acknowledgements}

\bibliographystyle{aa}
\bibliography{Noels}

\end{document}